\documentclass[journal]{IEEEtran}

\usepackage[utf8]{inputenc}
\usepackage[T1]{fontenc}
\usepackage{subfigure}
\usepackage{xcolor,colortbl}

\usepackage{multirow}
\usepackage{arydshln}
\usepackage{amsfonts}
\usepackage{enumitem}

\newcommand{\org}[1]{\overline{#1}} 
\newcommand{\II}{\mathcal{I}}
\renewcommand{\vec}[1]{\mathbf{#1}} 

\definecolor{Gray}{gray}{0.85}
\definecolor{mybg}{HTML}{d3d3c5}
\definecolor{mygreen}{HTML}{b0e7a6}
\definecolor{myblue}{HTML}{d5ebf5}
\definecolor{myyellow}{HTML}{e7dea6}
\definecolor{myred}{HTML}{e7b9a6}
\definecolor{mypink}{HTML}{e8dae0}

\newcommand{\todo}[1]{}
\newcommand{\rephrase}[1]{}

\newcommand{\hi}{{\text{high}}}
\newcommand{\lo}{{\text{low}}}

\newcolumntype{a}{>{\columncolor{Gray}}c}
\newcolumntype{b}{>{\columncolor{white}}c}

\newcommand{\Rr}{\mathbb R}

\def\thesubsubsectiondis{(\alph{subsubsection})}

\usepackage[cmex10]{amsmath}
\usepackage{mathtools}
\usepackage{multirow} 
\usepackage{graphicx}
\usepackage{hyperref}
\usepackage{rotating}
\usepackage{array}
\usepackage{adjustbox}
\usepackage{cancel}

\begin{document}
\title{MarmoNet: a pipeline for automated projection mapping of the common marmoset brain from  whole-brain serial two-photon tomography 
\begin{center} \small{Brain Image Analysis Unit}\\\small{RIKEN Center for Brain Science, Japan}\\\small{\url{http://bia.riken.jp}} 
\end{center}
}

\author {
Henrik Skibbe, Akiya Watakabe, Ken Nakae,
    Carlos Enrique Gutierrez,  Hiromichi Tsukada,
  \\
    Junichi Hata, Takashi Kawase, 
     Rui Gong, Alexander Woodward,
    
    Kenji Doya,
    Hideyuki Okano,
    Tetsuo Yamamori,
    Shin Ishii 
    \thanks{
    H. Skibbe is with the Brain Image Analysis Unit, RIKEN Center for Brain Science, Wako, Japan. (e-mail: henrik.skibbe@riken.jp).
    }
    \thanks{
    J. Hata and H. Okano are with the Laboratory for Marmoset Neural Architecture, RIKEN Center for Brain Science, Wako, Japan and the Department of Physiology, Keio University School of Medicine, Tokyo, Japan.
    }
    \thanks{
    C. E. Gutierrez, H. Tsukada and K. Doya are with the Neural Computation Unit, Okinawa Institute of Science and Technology, Okinawa, Japan.
    }
    \thanks{
    A. Watakabe and T. Yamamori are with the
    Laboratory for Molecular Analysis of Higher Brain Function, RIKEN Center for Brain Science, Wako, Japan.
    }
    \thanks{
    K. Nakae and S. Ishii are with the Department of Systems Science, Kyoto University, Kyoto, Japan. 
    }
     \thanks{
    R. Gong and A.Woodward  with the Connectome Analysis Unit, RIKEN Center for Brain Science, Wako, Japan.
    }
    \thanks{
    T. Kawase is with the Janelia Farm Research Campus in Ashburn, Virginia, USA.
    }
    \thanks{
    This research is supported by the program for Brain Mapping by Integrated Neurotechnologies for Disease Studies (Brain/MINDS) from Japan Agency for Medical Research and Development, AMED
    }

}

\markboth{Technical report, Brain Image Analysis Unit, RIKEN Center for Brain Science, Japan, 2019, v.1}%
{}

\maketitle

\begin{abstract}
Understanding the connectivity in the brain is an important prerequisite for understanding how the brain processes information. In the Brain/MINDS project, a connectivity study on marmoset brains uses two-photon microscopy fluorescence images of axonal projections to collect the neuron connectivity from defined brain regions at the mesoscopic scale. The processing of the images requires the detection and segmentation of the axonal tracer signal. The objective is to detect as much tracer signal as possible while not misclassifying other background structures as the signal. This can be challenging because of imaging noise, a cluttered image background, distortions or varying image contrast cause problems.

We are developing \textit{MarmoNet}, a pipeline that processes and analyzes tracer image data of the common marmoset brain. The pipeline incorporates state-of-the-art machine learning techniques based on artificial convolutional neural networks (CNN) and image registration techniques to extract and map all relevant information in a robust manner. The pipeline processes new images in a fully automated way.

This report introduces the current state of the tracer signal analysis part of the pipeline.
\end{abstract}

\def\thesubsubsectiondis{(\arabic{subsubsection})} 

\section{Introduction}
A major challenge of contemporary neuroscience research projects is the analysis of brain image data to better understand the brain structure and function \cite{markram2012human,oh2014mesoscale,grillner2016worldwide}. In particular, the primate brain is in focus of major national and international brain research projects \cite{markram2012human,okano2016brain,poo2016china}. Brain imaging data from primates such as macaque or marmoset monkeys can bridge the gap between the brain of simple vertebrate models, such as rodents, and the complexity of the human brain. 

A structural map of the brain is an indispensable prerequisite for studies of the neural networks controlling higher brain functions. The Brain/MINDS project in Japan conducts multimodal studies to construct an integrated structural map of the marmoset brain \cite{okano2016brain,okano2015brain2}. An ambitious goal of the project is to gain new insights into information processing and diseases of the primate brain.  As part of the Brain/MINDS project, we conduct a systematic tracer mapping using viral-based anterograde tracers.  

A large amount of tracer image data demands automated processing and analysis. To cope with this demand, we are developing MarmoNet, an image processing pipeline that automatically processes and analyses imaging data of the marmoset brain. 

\begin{figure}
\centering
\subfigure[ An adeno-associated virus bearing a gene of fluorescent protein has been injected to a defined region in the left frontal cortex to tag axonal projections. Following viral infection, the cells become filled with the fluorescent protein.\label{Fig:tracerintroa}]{\includegraphics[width = 1 \columnwidth]{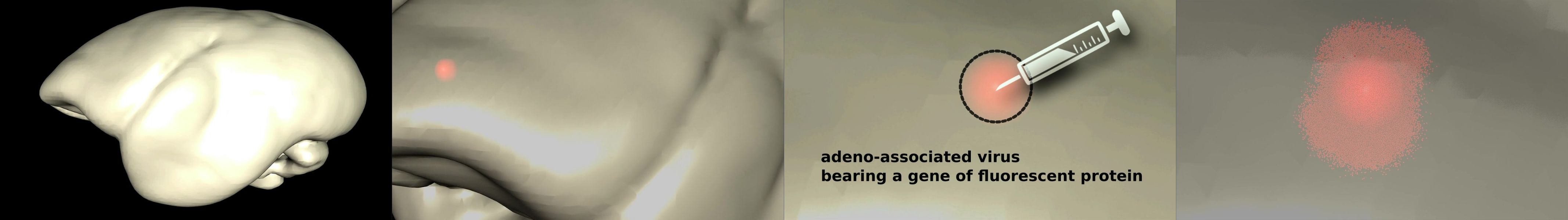}}
\subfigure[All fluorescent signal outside the infected area is considered to come from axons that had originated from cells located at the injection site (red area in the image). By connecting the injection site location with the axon tracer projection sites, we can make assumptions about brain connectivity.\label{Fig:tracerintrob}]{\includegraphics[width = 1 \columnwidth]{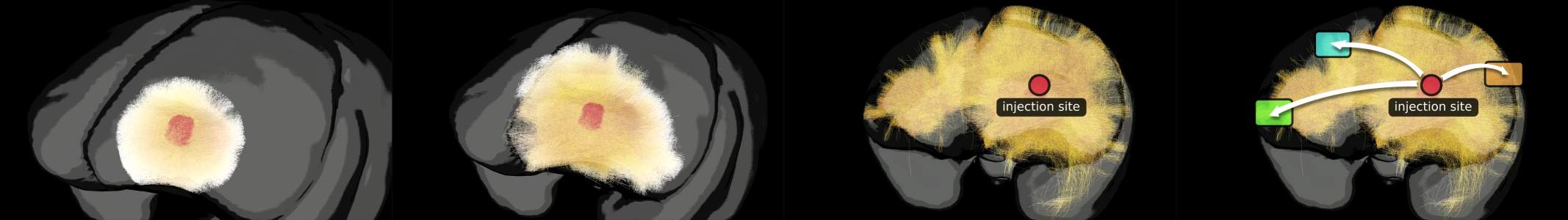}}
\subfigure[Injection locations in the prefrontal cortex.  By systematic injections of many tracers, we can construct a map of neural connections across the brain regions of interest.We exemplarily show the pattern  of 3D axonal projections corresponding to four different injection sites.  \label{Fig:tracerintroc}]{\includegraphics[width = 1 \columnwidth]{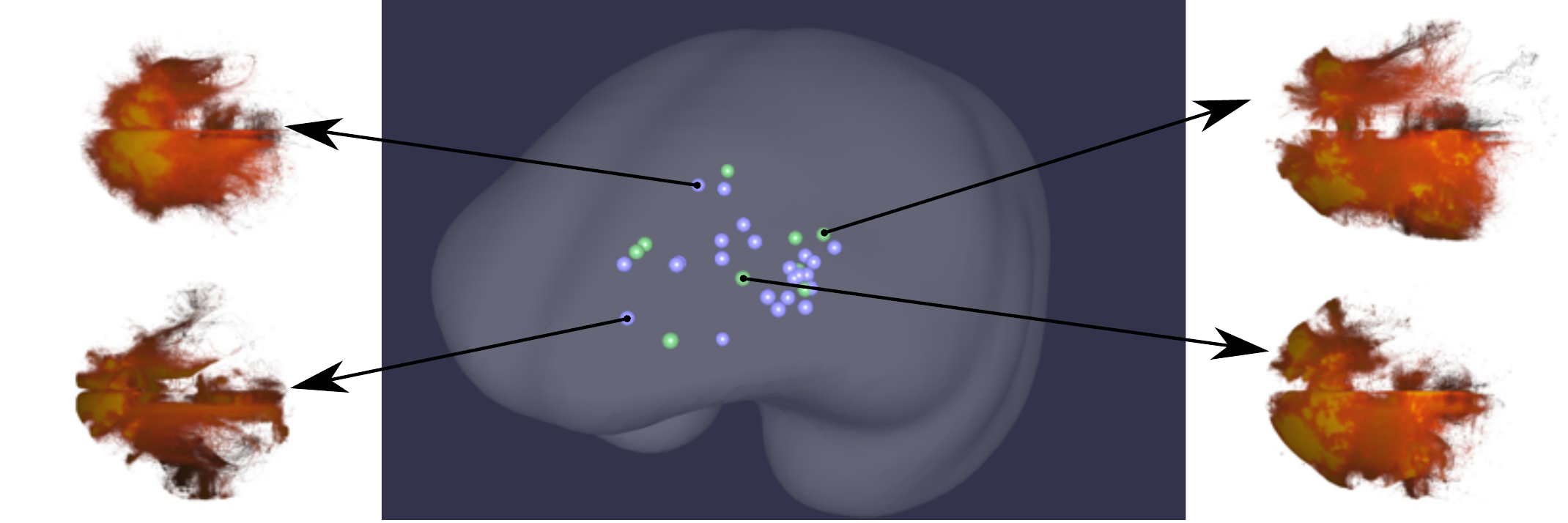}}
\caption{Anterograde tracer injection procedure for tagging axonal projections.}
\label{Fig:tracerintro}
\end{figure}

The goal in our anterograde tracer study is to track the fluorescent signal of the axons from their cell bodies to their point of termination. Figure \ref{Fig:tracerintro} illustrates the concept of the anterograde tracer technique that we use.  
To trace the axon projections of neurons located in a specific region of the brain, a virus bearing a gene of a fluorescent protein is injected into the region's center; see Fig. \ref{Fig:tracerintroa}. After viral infection, the neurons in that brain region become filled with the fluorescent protein. All fluorescent signal outside the infected area is considered to come from axons that had originated from cells located at the injection site; see Fig. \ref{Fig:tracerintrob}. By connecting the injection site location with the axon tracer projection sites, we can infer brain connectivity. By systematic injections of many tracers, we can construct a map of neural connections across the brain regions of interest; see Fig. \ref{Fig:tracerintroc}.

In recent years, several pipelines for the processing of neural anterograde tracer image data have been proposed \cite{kuan2015neuroinformatics,abe20173d,lin2019high}. 
Typically, hand-crafted heuristic features detect the tracer signal, and thresholds are being applied to segment the images into background and tracer signal. In \cite{kuan2015neuroinformatics} for example, edge enhancing filters have been applied to enhance the signal of axon structures. Then, morphological attributes have been extracted and classified to further improve the segmentation results. In \cite{abe20173d}, an independent component analysis was used to separate the tracer signal from the background.


It is challenging to describe all kinds of patterns of axon tracer signal within the images. The axons in the images appear in various shapes and with varying contrasts and intensities. Some axons, particularly when sparsely sampled and running within the imaging plane, appear as elongated, bright structures. However, some other densely sampled axons build dense, blob-like patterns with a texture that can vary depending on whether they cross and bifurcate or they share a dominant direction as  part of a thicker axon bundle. Hence it is oftentimes challenging to manually craft a feature extraction pipeline that can cope with all cases equally well so that no further manual interaction is necessary.  

MarmoNet uses U-Net \cite{ronneberger2015u} to deal with these problems. U-Net is an artificial convolutional neural network (CNN) architecture 
that integrates both the feature extraction and the classification into one common trainable framework. U-Net has shown excellent performance in various segmentation tasks of neurite structures from biomedical image data  \cite{falk2019u,kanaoka2019determ} and therefore was a nearby candidate for solving our tracer segmentation problem.

An advantage of integrating machine learning into the pipeline is that the feature extraction and classification can be constructed in a data-driven way. This is done by providing a set of representative examples to train an U-Net to solve the task. A single example is a pair of an input image and its corresponding segmentation result. The machine then learns to extract the features and learns the classification parameters that are best suited to generate the segmentation for an input image.

\todo{generation of training data. Here two approaches: manual and semi-automated.}

\begin{figure}
\includegraphics[width = 1 \columnwidth]{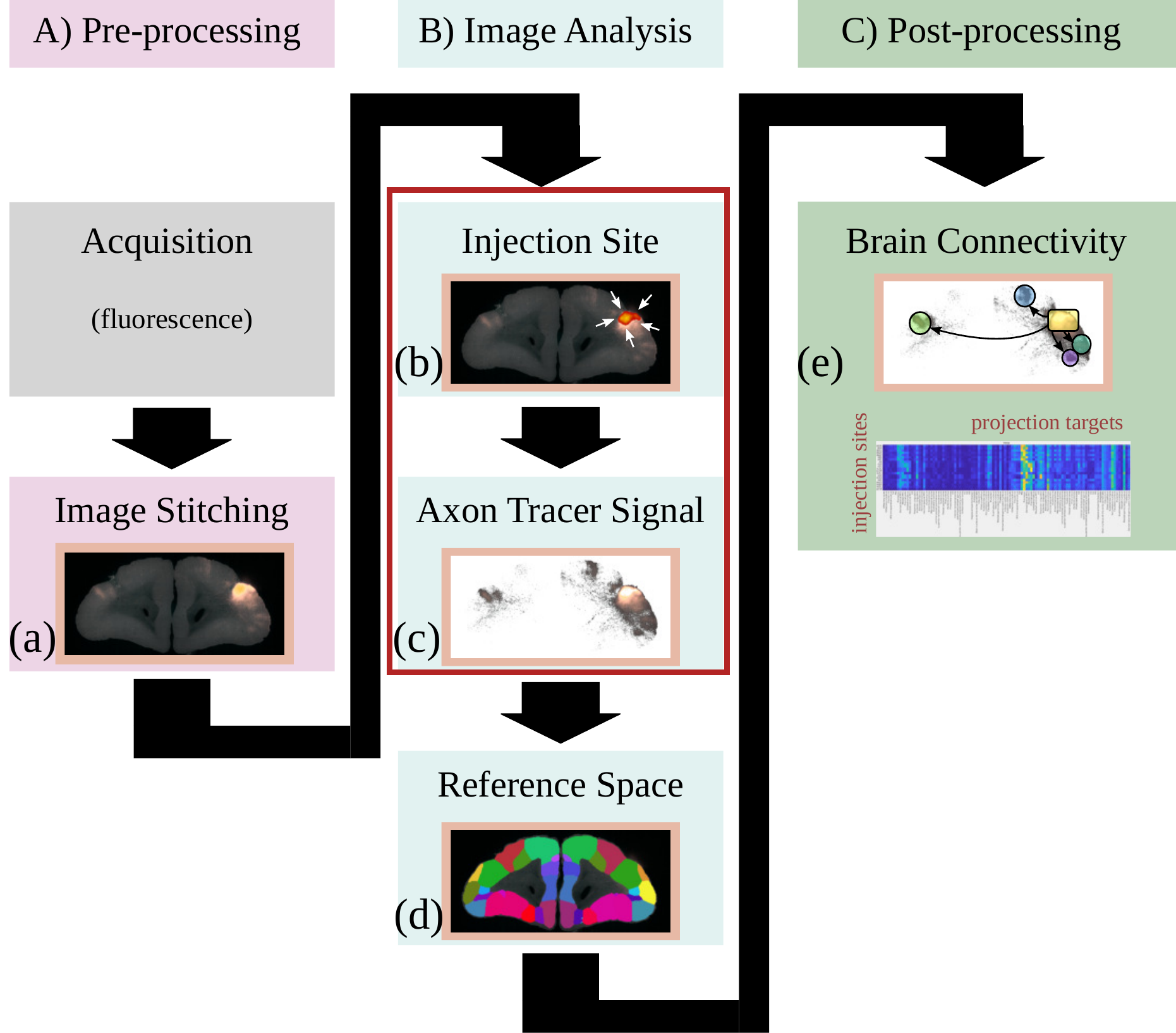}
\caption{MarmoNet processing pipeline. In this report, we focus on the tracer segmentation parts (b) and (c).}
\label{Fig::pipeline}
\end{figure}

\begin{figure*}
\includegraphics[width = 1 \textwidth]{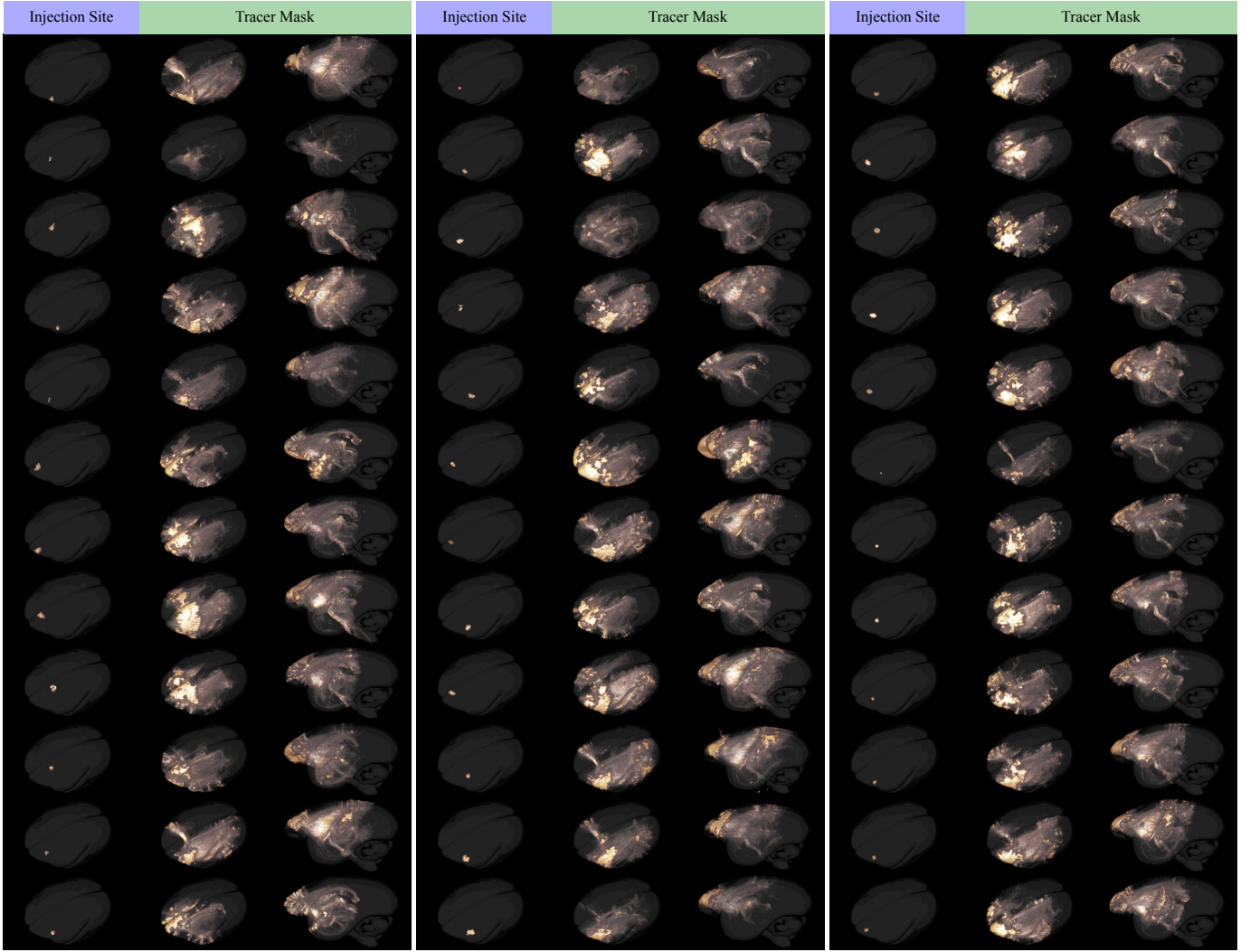}
\caption{3D image stacks of the detected injection site (first column, superior view) and the segmented axon tracer signal (2nd column superior view and 3rd column lateral view) for 36 injection sites. All images have been mapped to the Brain/MINDS reference space.}
\label{fig::results01}
\end{figure*}

As the time of writing, the construction of our marmoset connectivity map is an ongoing project.  We frequently presented our progress at conferences during the last years \cite{ABSTR_08,ABSTR_07,ABSTR_01, ABSTR_04, ABSTR_20,ABSTR_02, ABSTR_15, ABSTR_18, ABSTR_19}. Figure \ref{Fig::pipeline} shows an outline of the entire tracer data processing and analysis pipeline, including the pre-processing and the post-processing that generates the output.
The pipeline automatically detects the location of the injection site (connections origin), and the axonal projections (connection targets). The pipeline maps all image data into a common image reference space which allows the integration of connectivity data from different individuals. Figure \ref{fig::results01} shows the result for 36 examples.

This report focuses on the tracer signal segmentation part including the training and application of the U-Net to process and analyse the data. The tracer segmentation can be divided into two major steps (red box around those steps in Fig. \ref{Fig::pipeline}). The first step is the automatic determination of the location and boundary of the injection site (b), and  the segmentation of the axon tracer signal outside the injection site (c).

\section{Image Acquisition}

\begin{figure}
\centering
\includegraphics[width = 1 \columnwidth]{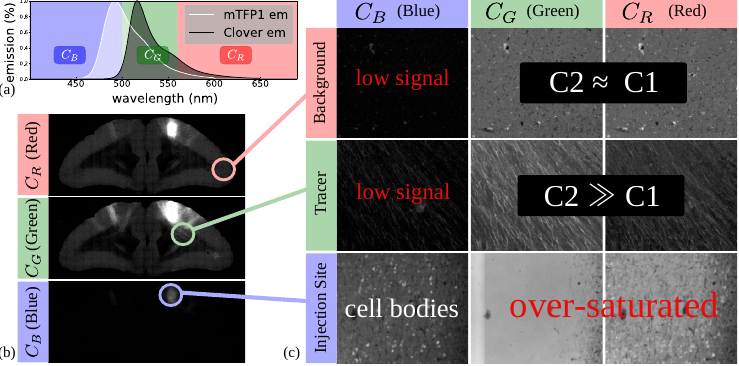}
\caption{Emitted fluorescent tracer signal and its application for  identifying the cell body signal, axon tracer signal and the  auto-fluorescent background signal. The channels $C_R$, $C_G$ and $C_B$ are representing the three imaging channels of the TissuCyte microscope. 
}
\label{Fig:microscope}
\end{figure}

Individual brain images have been acquired with one of our TissueCyte 1000 and TissueCyte 1100 (TissueVision, Cambridge, MA) serial two-photon tomographs (two-photon microscopes), 
which can generate optical section images from the block face surface of agarose-embedded biological material in coordination with a computer-driven microtome and stage movement \cite{ragan2012serial}. We target the neurons with a mixture of clover and 1/4 amount of presynapse targeting mTFP1 \todo{Akiya, 2 lines, please}. Figure \ref{Fig:microscope} (a) illustrates the emitted wavelengths of the fluorescent signal as an overlay of a bar diagram representing the three filters of the tissue microscope. The bar diagram represents the wavelengths captured by the three color channels: red ($C_R$), green ($C_G$) and blue ($C_B$).

Figure \ref{Fig:microscope} (c) illustrates the key features of all three channels. The autofluorescent signal is similarly strong in channels $C_R$ and $C_G$. The tracer signal, however, is much more stronger in $C_G$ than in channel $C_R$. This makes channel $C_R$ an ideal candidate for background signal removal. In the injection site, we have the best contrast in channel $C_B$. According to their characteristics, we will call $C_R$ the \textit{background} channel (for background removal), $C_G$ the \textit{tracer} channel (axonal tracer signal best visible), and $C_B$ the \textit{injection site} channel (best image of cell bodies).

\def\thesubsubsectiondis{(\alph{subsubsection})} 
\section{MarmoNet Pipeline}
The pipeline comprises three major components. The pre-processing part that performs the image stitching, the actual image analysis part, and a pipeline output part where connectivity calculations are being carried out.  Figure \ref{Fig::pipeline} illustrates the workflow of the pipeline. The single steps are:

\begin{enumerate}[label=(\alph*)]
 \item Applying an intensity correction that reduces inhomogeneous illumination effects and stitching the images.
 \item A CNN detects the location of the injection site.
 \item A CNN segments the axon tracer signal outside the injection site from the background.
 \item An image registration algorithm maps all image data to the reference image space of the Brain/MINDS marmoset brain atlas. 
  \item Connectivity calculations are being carried out. 
\end{enumerate}

All pipeline steps will be described in the remainder of this section, the tracer segmentation parts will be described in more detail.

\subsection{Pre-Processing}

\bigskip
\subsubsection{Image stitching}

The TissueCyte microscope acquires an image stack of the entire marmoset brain in a fully automated manner. Starting with the prefrontal cortex, the brain is imaged every $50 \mu m$ from its anterior part to its posterior regions (front to back).

 The TissueCyte microscope scans the entire surface of the block-face in a zigzag pattern using a movable table while generating a new 2D image tile $I^j_k$ every 750$\mu m$, where $j$ is the tile index, and $k$ denotes one of the three color channels $C_R$, $C_G$ or $C_B$.  The size of each image tile has been set to $720 \times 720$ pixels with a spatial in-plane resolution of $1.34 \times 1.34\mu m^2$.

 \smallskip
 \noindent
 \textbf{Intensity Correction}: The illumination of single image tiles is inhomogeneous due to distortions in the optical path. Typically, the intensity decreases in a nonlinear manner and is most bright in the image center, while getting darker towards the image boundaries. This visual effect is called \textit{vignetting}. The vignetting effect differs from microscope to microscopes such as for our TC1000 and our TC1100. The different pathways of the various color channels cause different effects as well. The distortion is often modeled as a linear modulation 
 \begin{align}
   {I^j_k}(x,y) = \org{{I^j_k}}(x,y) \hat{F_k}(x,y);  
   \label{eq::ff}
 \end{align}
 see e.g.  \cite{kask2016flat,likar2000retrospective}. The ${I^j_k}$ are the observed image tiles, $\org{{I^j_k}}$ the undistorted image, and $\hat{F_k}$ the \textit{shading field},  which is the image of the vignetting effect in channel $k$. We assume that the $\hat{F_k}$ are constant throughout an entire marmoset brain image stack.
 
 The technique to correct this distortion is called flat-field correction. The goal is to estimate the shading field $\hat{F_k}$ so that we can recover the original image tile $\org{{I^j_k}}$
 by inverting \eqref{eq::ff}.
 
 Similar to \cite{allentecrep}, we use a data-driven heuristic to estimate the shading field $\hat{F_k}$ by averaging over all image tiles of an entire marmoset brain image stack. Averaging over a sufficiently large number of image tiles $I^j_k$ yields an estimate of the shading field $F_k$ (up to a constant factor). In contrast to \cite{allentecrep}, we exclude outliers from the average. Outliers are dark pixels with no information and bright pixels, most likely, can be attributed to tracer signal. We describe the details of our flat-field correction in section \ref{sec::flatfieldcorr} in the appendix. Figure \ref{Fig::FFC01}  and Fig. \ref{Fig::FFC02} of the appendix are show results of our flat-field correction.

It is worth noting that we only perform an intensity correction for tile images of the channels $C_R$ (background) and $C_G$ (tracer). Channel $C_B$ (Injection site) has a clear signal around the injection site which is an insufficiently small amount of imaging data for estimating the shading field.

  \smallskip
 \noindent
 \textbf{Image Stitching}: 
  The TissueCyte microscope provides the offset for each image tile in 3D world coordinates with micrometer resolution.  
  The microscope generates optical section images from the block face surface before a built-in vibrating blade microtome mechanically cuts off the most upper tissue section. There is no spatial distortion within an image slice.
  Hence an entire image slice can be reconstructed by aligning and fusing all image tiles $I^j_k$ according to their world coordinates. In our settings, there is a small overlap between adjacent image tiles (about 80 pixels). We crop 50 pixels from the image tile boundaries and fuse the remaining overlapping parts using a linear blending function (similar to \cite{allentecrep}).

 \smallskip
 \noindent
 \textbf{Full resolution image stacks}: 
 We reconstruct the entire brain image stack in full resolution. Each slice of the image stack that corresponds to a physical brain section has a spatial resolution of $1.34 \times 1.34 \mu m^2$ pixels.  
The spatial distance between two image slices is $50 \mu m$.  We denote the entire 3D image stacks of channel $C_{k}$ with a spatial resolution of  $1.34 \times 1.34 \times 50 \mu m^3$ by $\II^\hi_{k}$. 
Each image stack is a function 
\begin{align}
\II^{\hi}_{k}:\Omega\to R_{\geq 0}, 
\end{align}
where $\Omega$ defines the image domain. We address single voxels in $\II^{\hi}_{k}$ via $\II^{\hi}_{k}(x,y,z)$, and entire 2D sections with $\II^{\hi}_{k}(z)$. The index $k$ defines the channel and $z$ the section number that corresponds to a physical section in the TissueCyte microscopy data.

\smallskip
 \noindent
 \textbf{Low resolution image stacks}: In addition, we scale the full resolution image stacks $\II^{\hi}_k$ down to lower resolution image stacks $\II^{\lo}_k$ with an isotropic voxel resolution of $50^3 \mu m^3$. The isotropic image stacks are used for image registration, injection site localization and connectivity calulations.

\begin{figure}
\centering
\subfigure[before]{
\includegraphics[width = 0.47 \columnwidth]{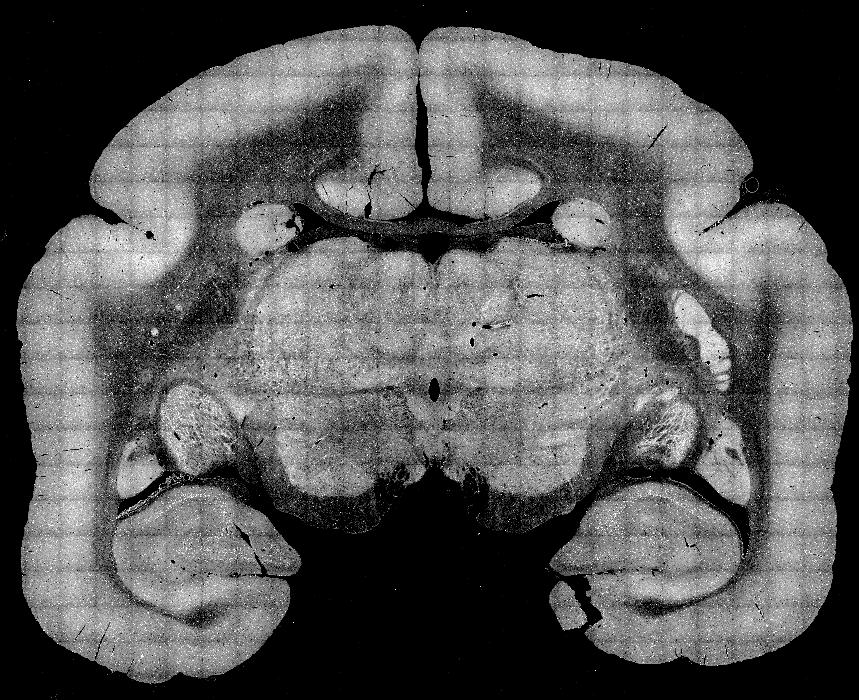}}
\subfigure[after]{
\includegraphics[width = 0.47 \columnwidth]{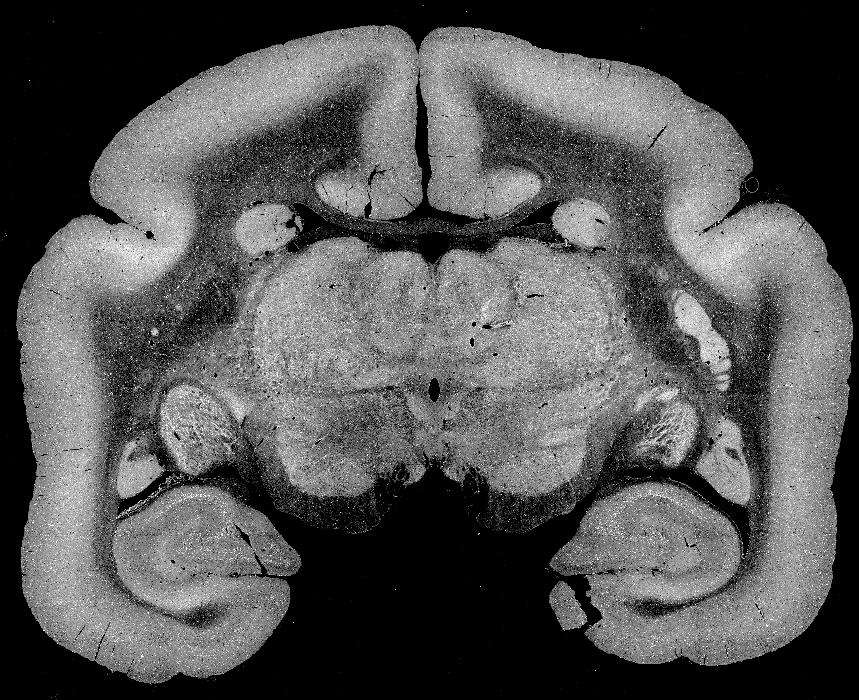}}
\subfigure[before]{
\includegraphics[width = 0.47 \columnwidth]{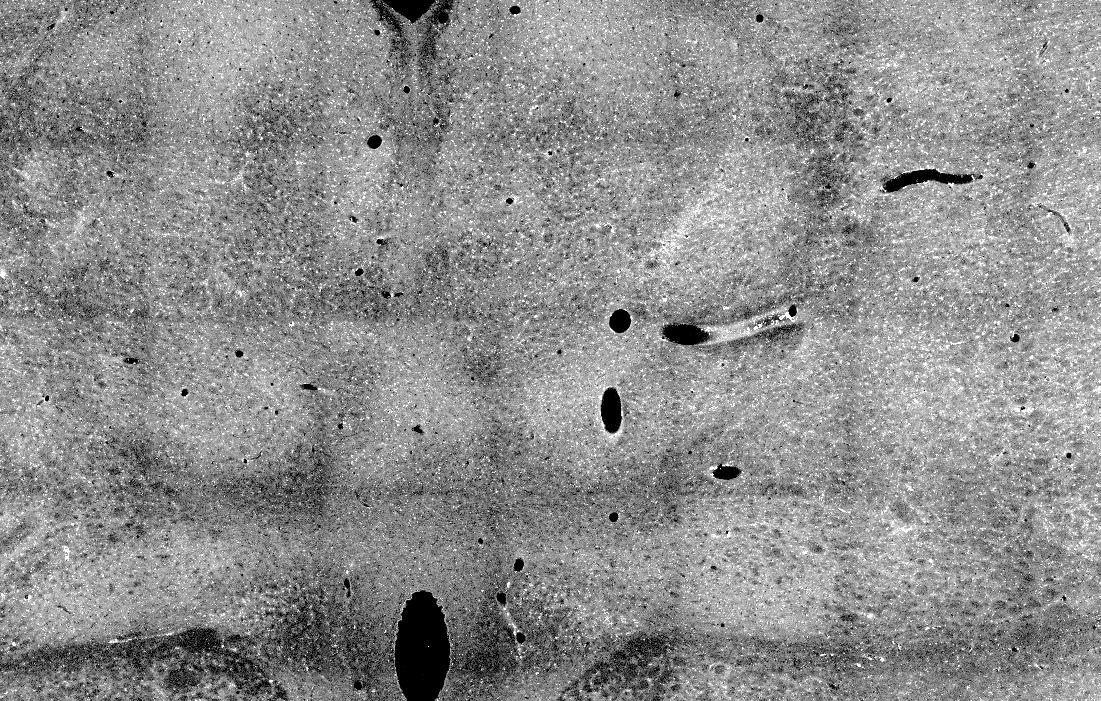}}
\subfigure[after]{
\includegraphics[width = 0.47 \columnwidth]{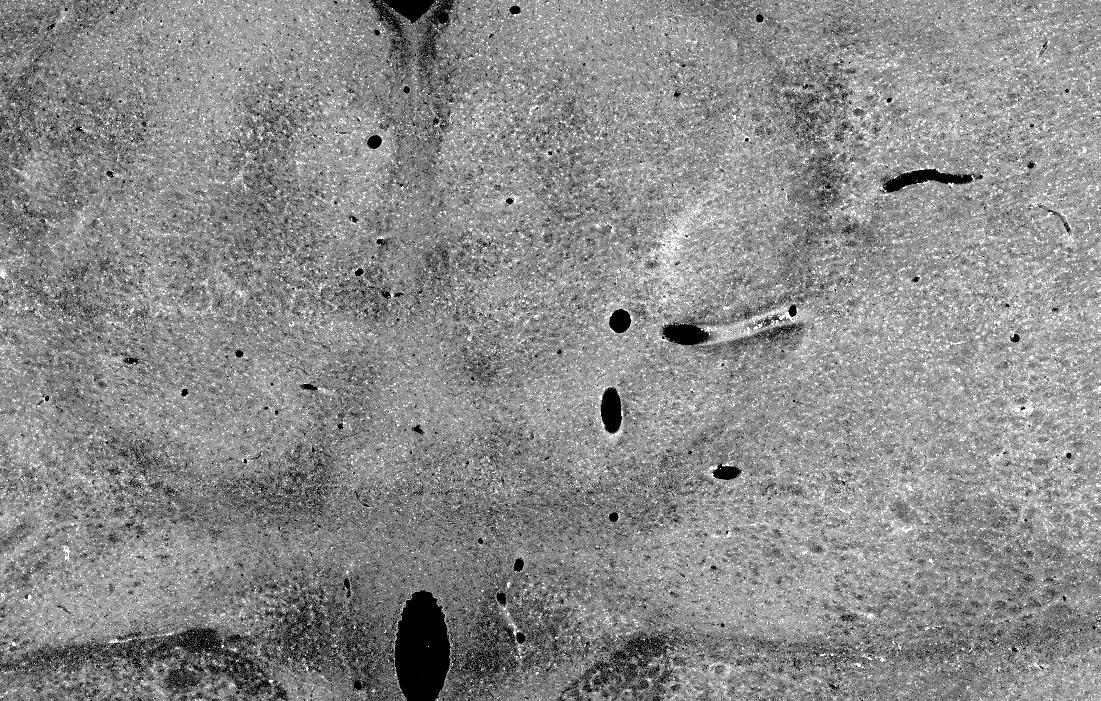}}
\caption{The flatfield correction noticeably reduces the vignetting effect.}
\label{Fig::FFC01}
\end{figure}

\begin{figure}
\centering
\includegraphics[width = 0.45 \columnwidth]{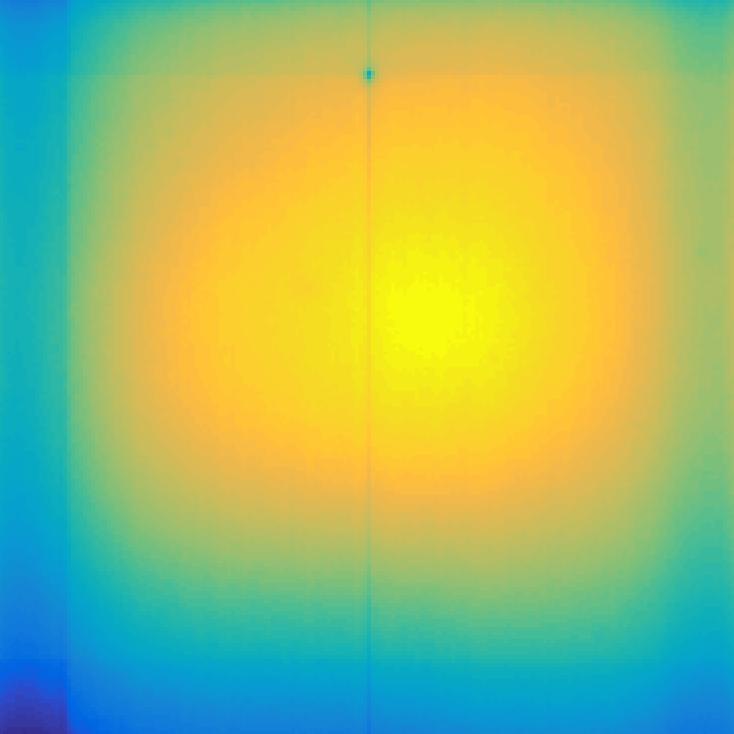}
\caption{Example of an estimated shading field.}
\label{Fig::BF}
\end{figure}

\subsection{Image Analysis}
\setcounter{subsubsection}{1}

The extraction of brain connectivity information from the tracer data demands the segmentation of the tracer signal from the image background. The pipeline detects the tracer signal in two steps. In the first step, it detects the signal of the cell bodies in the injection site. In the second step, it segments the axonal tracer signal in the remaining parts of the brain image. After tracer segmentation, the image stacks are mapped to a common population average brain reference image space to integrate brain  connectivity data from different individuals.

\subsubsection{Injection site localization}\label{sec::injectionsite}

%

\begin{figure}
\centering
\includegraphics[width = 1 \columnwidth]{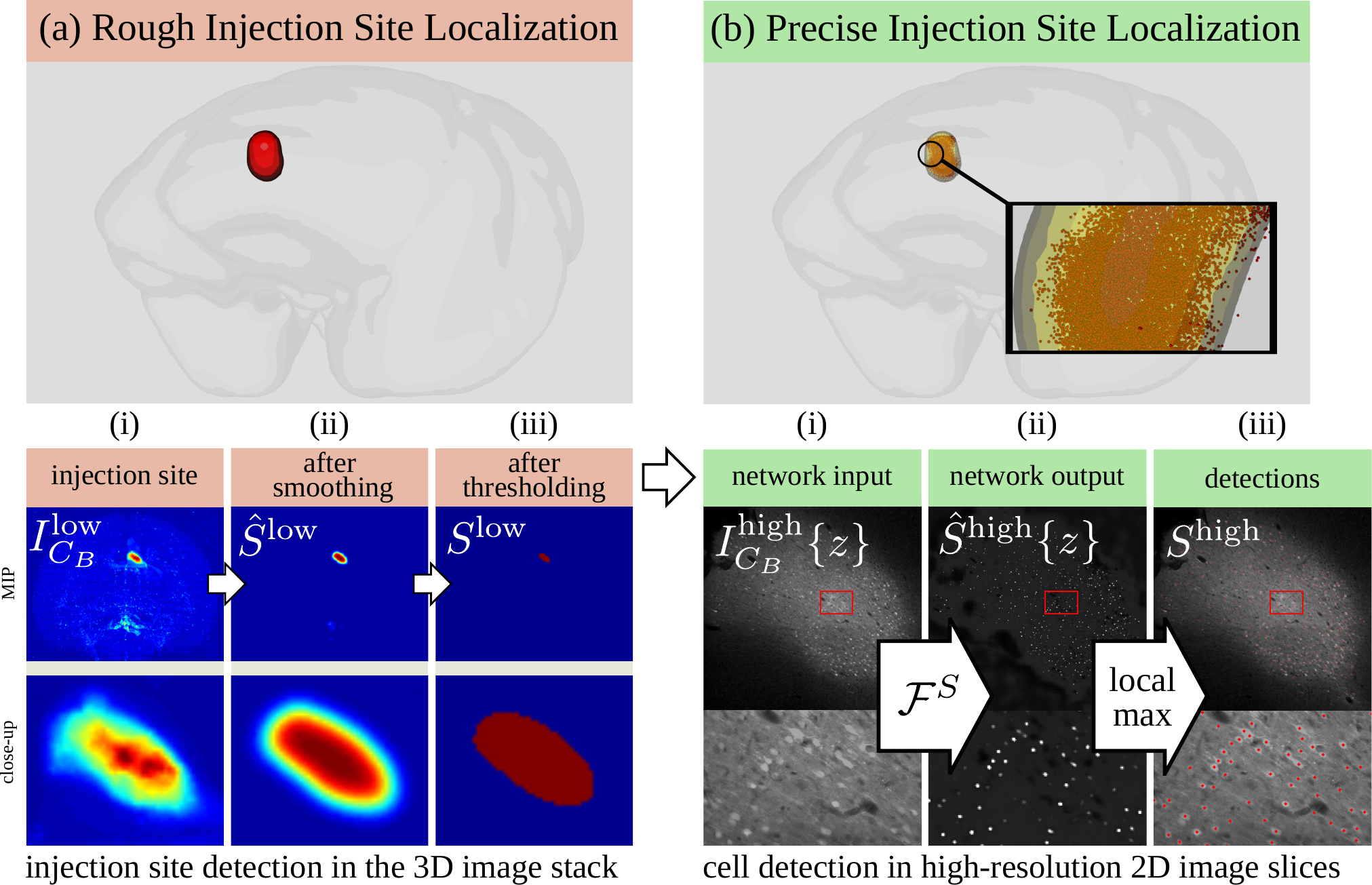}
\caption{Injection site localization pipeline.}
\label{Fig::injloc}
\end{figure}

The injection site location is determined in two steps. First, a heuristic approach roughly estimates the volume of the injection site region in the low-resolution image of the injection site channel $C_B$. Further analysis is then limited to that volume to reduce computation costs.
The second step identifies the location of the infected cell bodies in the injection site. A CNN determines the cell body locations in the high-resolution image section of the injection site. The cell body positions define a 3D point cloud that implicitly determines the exact injection site boundary. The procedure is illustrated in Fig. \ref{Fig::injloc}.

\smallskip

\begin{enumerate}
\item\textbf{Rough Localization:} This step is illustrated in Fig. \ref{Fig::injloc} (a). The injection site appears significantly brighter than the remaining tissue regions in channel $C_B$. We use this feature to roughly estimate the injection site location and boundary. In this step, all calculations are performed  on the low-resolution 3D image stack $\II^\lo_{C_B}$. We first classify the image into background voxels and injection site candidate voxels using a constant intensity threshold. As threshold we determined 4500 in a heuristic manner. \rephrase{Then, we count the positively classified voxels in a local neighborhood by convolving the thresholded image with an isotropic Gaussian function.} The width of the Gaussian is $\sigma = 150 \mu m$ (3 voxels). Typically, most of the voxels in the injection site have been classified as tracer signal so that after convolution, the injection site appears as a homogeneous, bright structure in the image. We denote this image by $\hat{S}^\lo$.  Figure \ref{Fig::injloc} (a.ii) shows an example. We obtain the final injection site mask $S^\lo$ by thresholding $\hat{S}^\lo$ a second time.  The second threshold is determined dynamically. It has been set to $t_\text{low}=0.5\max(\hat{S}^\lo)$,  half the maximum value  of $\hat{S}^\lo$. 
  The thresholded image is defined by 
  \begin{align}
 {S}^\lo(x,y,z) = \begin{cases}
                   1 & \text{if }  \hat{S}^\lo(x,y,z) > t_\text{low} \\
                  0 & \text{else}
                  \end{cases}.
\end{align}
There can be more than one connected regions in ${S}^\lo$. We used a connected component algorithm to remove all but the largest connected region in ${S}^\lo$. 

Figure \ref{Fig::injloc} (a) shows the maximum intensity projection of a raw image $\II^\lo_{C_B}$ of channel $C_B$, the maximum intensity projection of $\hat{S}^\lo$, and a maximum intensity projection of the final mask ${S}^\lo$.

\bigskip
\item\textbf{Precise Localization:} This step is illustrated in Fig. \ref{Fig::injloc} (b). We use the rough localization of the injection site in the low-resolution image $\II^\lo_{C_B}$ to determine the corresponding region of interest in the full resolution image stack $\II^\hi_{C_B}$. Then we determine the precise location of infected cells in the full resolution 2D image sections $\II^\hi_{C_B}(z)$. Sub image (i) in Fig. \ref{Fig::injloc} (b) shows an example image.

The pipeline uses a CNN to detect the locations of the cell bodies in the 2D image section $\II^\hi_{C_B}(z)$. The network is applied to all 2D slices that intersect with the segmentation mask $S$. 

The pipeline uses a U‐Net \cite{ronneberger2015u} for cell detection, a widely used CNN architecture for image segmentation tasks. The network  has been trained to mimic a function $\mathcal{F}^S$ that takes an image slice $\II^\hi_{C_B}(z)$ as input and maps it to a cell saliency image $\hat{S}^\hi(z):\Omega \to [0,1]$. The extents of $\hat{S}^\hi(z)$ are identical to $\II^\hi_{C_B}(z)$. The values in  $\hat{S}^\hi(z)$ correspond to the likeliness of cell body locations. As brighter a pixel in $\hat{S}^\hi(z)$, the more likely there is a cell at the corresponding position in the image slice $\II^\hi_{C_B}(z)$. Figure  \ref{Fig::injloc} (b.ii) shows an example. In a post-processing, we only keep the positions of local maxima in the image plane of $\hat{S}^\hi(z)$ that exceed a given threshold $t_\text{high} >0$. The final set ${S}^\hi$ of detected cell positions is defined by the 3D point cloud
\begin{align}
{S}^\hi:=&\{\forall \vec (x,y,z)| \nonumber\\
&\hat{S}^\hi(x,y,z) \text{ is local maxima in }\hat{S}^\hi(z) \nonumber\\
&\wedge\hat{S}^\hi(x,y,z)>t_\text{high} \label{eq::cellmax}
\},
\end{align}
where the threshold $t_\text{high}$ is equal to 0.5. Figure  \ref{Fig::injloc} (b.iii) shows the detected cell positions in the corresponding image slice.

Details of the network architecture and the training are explained in the next section of this report.

\end{enumerate}

\bigskip

\subsubsection{Axon tracer signal detection}

\begin{figure}
\centering
\includegraphics[width = 1 \columnwidth]{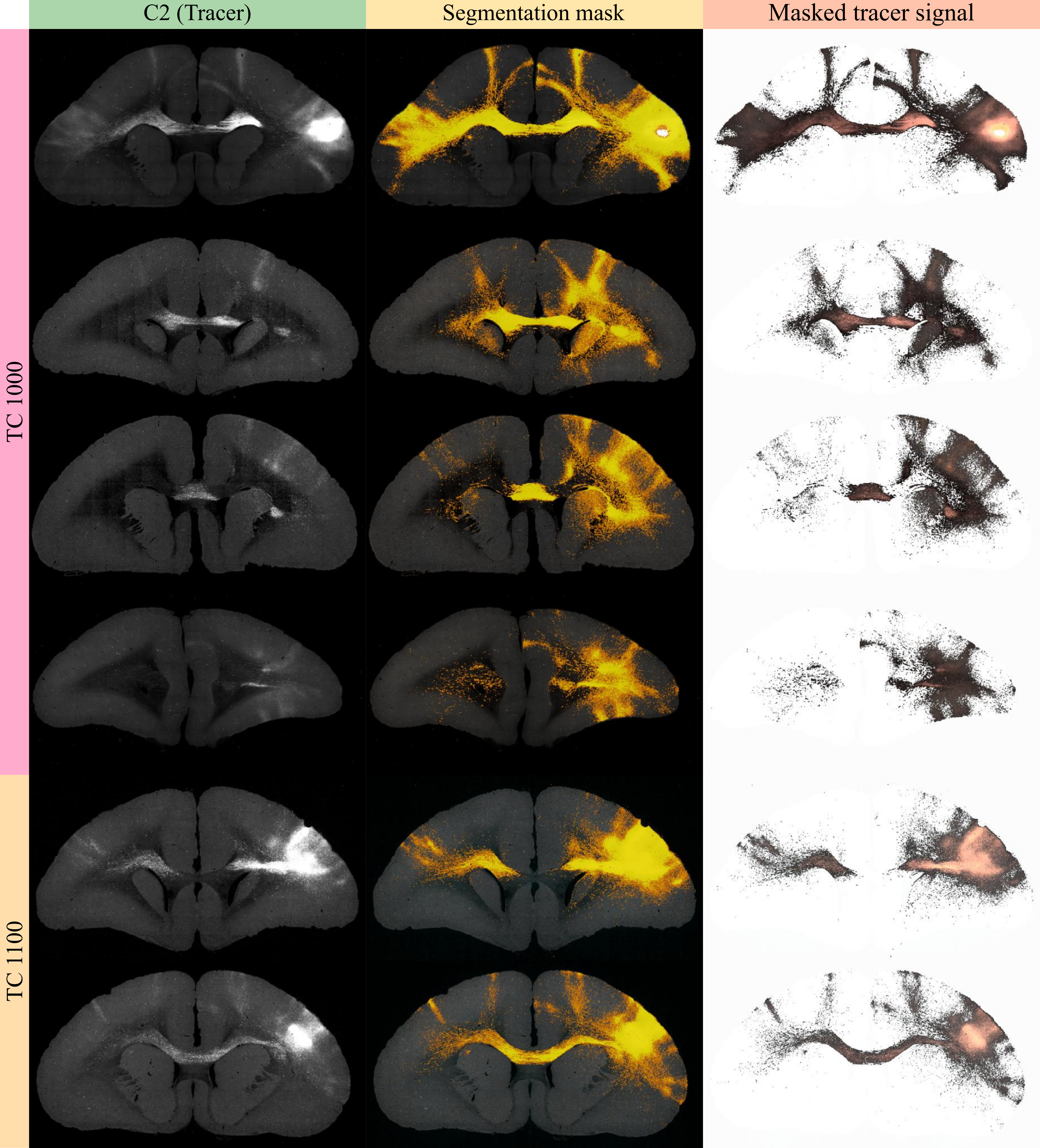}
\caption{Segmented tracer signal from image slices of five different marmoset subjects  acquired with different microscopes.
}
\label{fig::NNtracerexamples}
\end{figure}

The detection of axons can be quite challenging. The axons in the images appear in various shapes and with varying contrasts and intensities. Some axons, particularly when sparsely sampled and running within the imaging plane, appear as elongated, bright structures.  Some other densely sampled axons build dense, blob-like patterns with a texture that can vary depending on whether they cross and bifurcate or they share a dominant direction as  part of a thicker axon bundle.

Further, in some images, there are various non-axonal structures, like the boundary of some blood vessels, that, locally, appear similar to weak axon tracer signal in the channels $C_R$ and $C_G$, which makes them even harder to distinguish.

We experimented with intensity thresholding or edge detection approaches. While we could achieve good results for many individual image slices, it was impractical to find good parameters that work well for all images. The choice for parameters was always a trade-off between too many false positive detections or too many false negative detections. Section \ref{sec::training::tracer} shows some examples for the thresholding approach. We finally made use of a machine learning approach that can cope with the variation in the data.

\begin{figure}[ht]
\includegraphics[width = 1 \columnwidth]{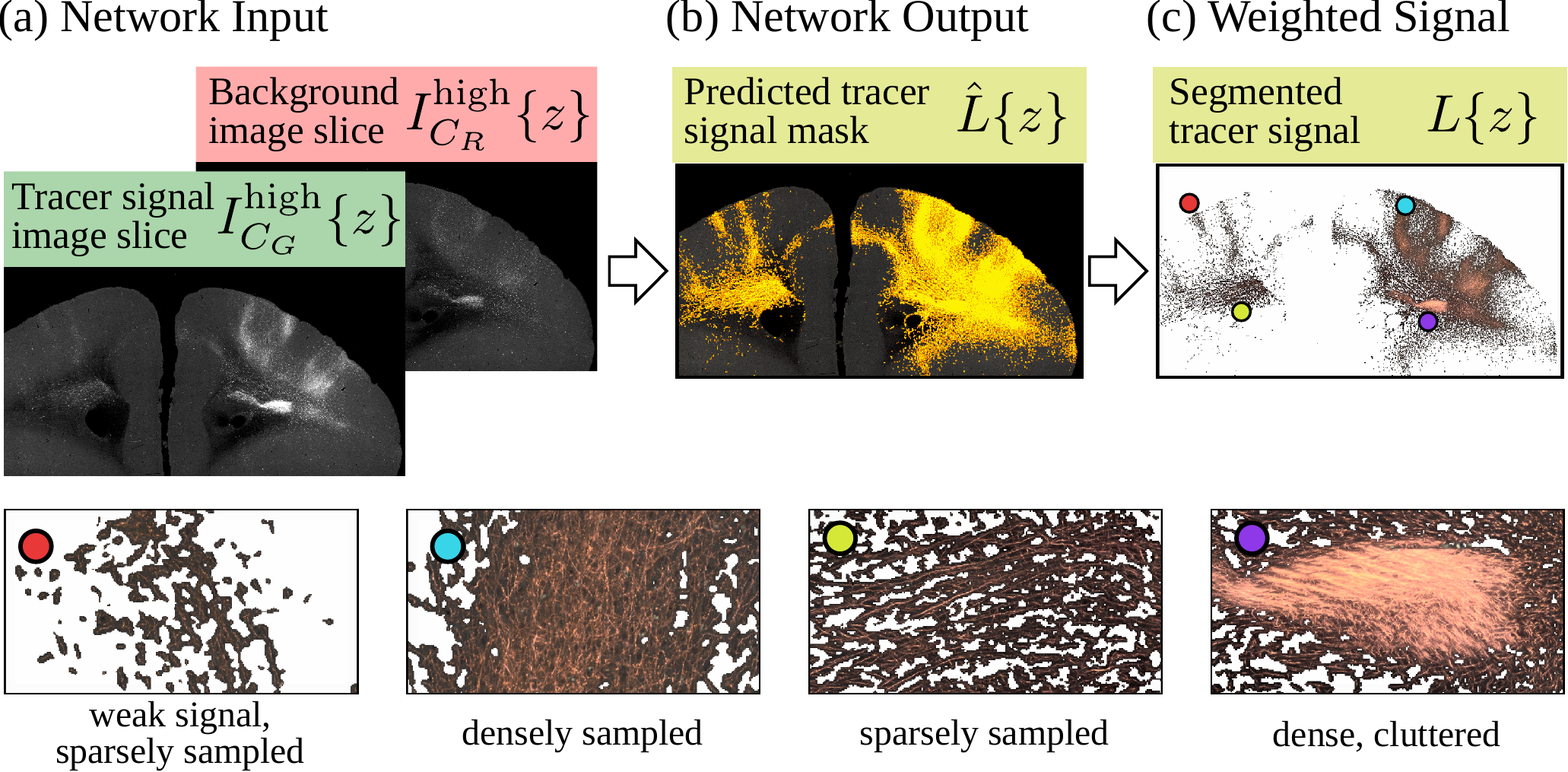}
\caption{Axon tracer segmentation pipeline.}
\label{fig::tracerseg}
\end{figure}

Similar to the cell detection task in the previous section, we use a CNN based on the U-net architecture to perform the tracer signal segmentation. The network architecture is mostly identical to the network that we use for cell detection. Only the image input dimension is different. Figure  \ref{fig::tracerseg} shows an outline of the axon tracer segmentation pipeline.

The trained network is a function $\mathcal{F}^L$  that takes both an image $\II^\hi_{C_R}(z)$ of the background channel $C_R$ and an image $\II^\hi_{C_G}(z)$ of the tracer channel $C_G$ as inputs (Fig. \ref{fig::tracerseg} (a)), and maps them to an axon tracer signal saliency image $\hat{L}^\hi(z):\Omega \to [0,1]$; see Fig. \ref{fig::tracerseg} (b). As brighter a value in $\hat{L}^\hi(z)$, the more likely there is an axon at the corresponding position in the image slices $\II^\hi_{k}(z)$.

The intensity correlates strongly with the axon density. This is an important feature that we believe is correlated with the connection strength between the injection site and the projection targets of the axons. To reflect this important feature in the signal segmentation, we first compute the signal strength by subtracting the background channel from the tracer signal channel (removing the autofluorescence signal), where 
\begin{align}
T\{t\}(x,x,z) =  &\begin{cases}
                   0 \text{ \textit{if} } \II^\hi_{C_G}(x,y,z)<t \II^\hi_{C_R}(x,y.z)  \\
                   \II^\hi_{C_G}(x,y,z))-t \II^\hi_{C_R}(x,y,z) \text{ \textit{otherwise}} \nonumber
                  \end{cases}
                  \label{eq::tracerthreshold}
\end{align}
is the tracer signal after background subtraction. The threshold $t$ has been set to $1.1$. We then threshold the saliency map $\hat{L}(z)$, which is the output of the neural network, to obtain a mask that removes all remaining non-axonal structures. Finally, we weight the  tracer signal with the raw tracer signal $T(z)$ according to
\begin{align}
L(x,y,z):=(\hat{L}(x,y,z)>0.5) T\{1.1\}(x,y,z).
\end{align}
The image $L$ is the segmented tracer signal. The scaling factor $1.1$ has been determined experimentaly by comparing  tracerless background image tiles of channel $C_R$ and $C_G$. Figure \ref{fig::tracerseg} (c) shows an example slice. Other examples can be found in Fig. \ref{fig::NNtracerexamples}.

\subsubsection{Mapping to reference space} 

\begin{figure}[ht]
\centering
\includegraphics[width=1 \columnwidth]{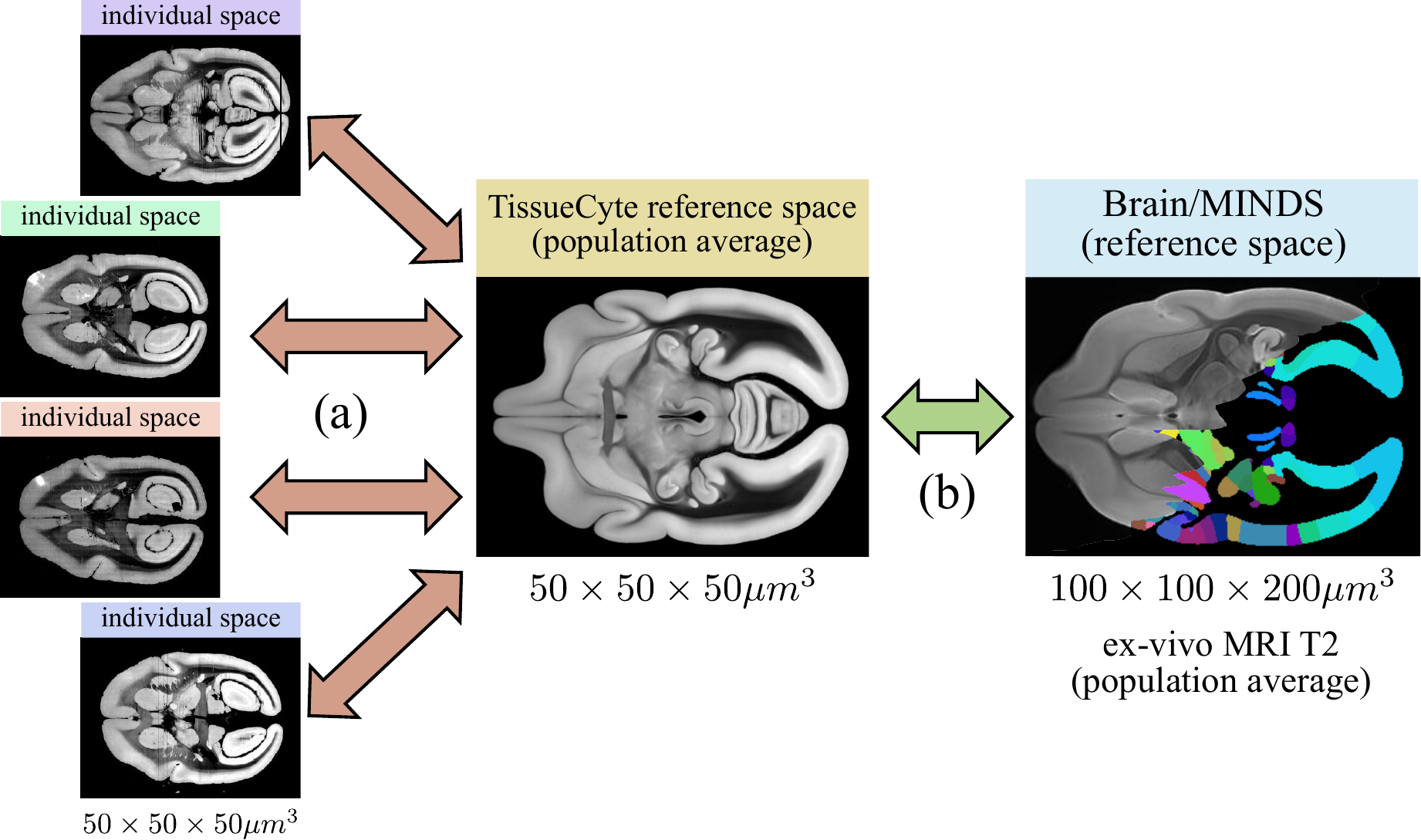}
\caption{The TissueCyte population average serves as a proxy to map individual brain images to the Brain/MINDS reference image space.}
\label{fig::mapping}
\end{figure}

After pre-processing and signal analysis, the image data is mapped to the Brain/MINDS reference image space. The mapping to a common reference space is a required prerequisite for the comparison and integration of imaging data from different individuals. In the Brain/MINDS project, a refined version of the Brain/MINDS marmoset brain atlas \cite{woodward2018brain} defines the common space that connects the imaging data of different modalities and data from different laboratories. The refined version of Brain/MINDS marmoset brain atlas defines 786 brain regions per hemisphere. The atlas is in alignment with a brain reference image space based on a T2 weighted MRI population average brain image. The spatial resolution of the T2 weighted MRI image is $100^3 \mu m^3$. 

All images, including the raw images, but also the injection site and tracer labels are in the same image space as the images of channel ${C_R}$. 
In a first step, we determine the mapping between the background image $C_R$ and the T2 image of the reference brain using the ANTs registration toolkit \cite{avants2011reproducible}. ANTs is that provides the mapping from the individual image space to the reference image space and its inverse mapping. Once we have this mapping, we can apply this mapping to all remaining images to map them from the individual image space to the reference space and vice versa.

In a preliminary version of the pipeline, we directly mapped the 3D image stacks of the background channel $C_R$ to the reference brain image. There are two drawbacks with such approach. First, the spatial resolution of the target image is lower than the $50^3 \mu m^3$ resolution of the isotropic TissuCyte image stacks. Further, the modalities differ. Both make an exact mapping challenging. While working fine for many, some images frequently made problems. 

We decided to put a TissuCyte average brain image as a proxy in between the TissuCyte images of an individual marmoset and the Brain/MINDS marmoset brain atlas. Figure \ref{fig::mapping} illustrates the procedure. Due to the same contrast and same spatial resolution, the registration between two TissuCyte images works better than the registration between two different modalities. In this new setting, the more challenging registration between the Brain/MINDS marmoset brain atlas and the  TissuCyte average image has to be determined only once. This has the advantage that even semi-automated or even manual approaches can be applied to improve the registration accuracy between the TissueCyte population average and the T2 MRI reference brain image.

\begin{figure}[ht]
\centering
\includegraphics[width=0.9 \columnwidth]{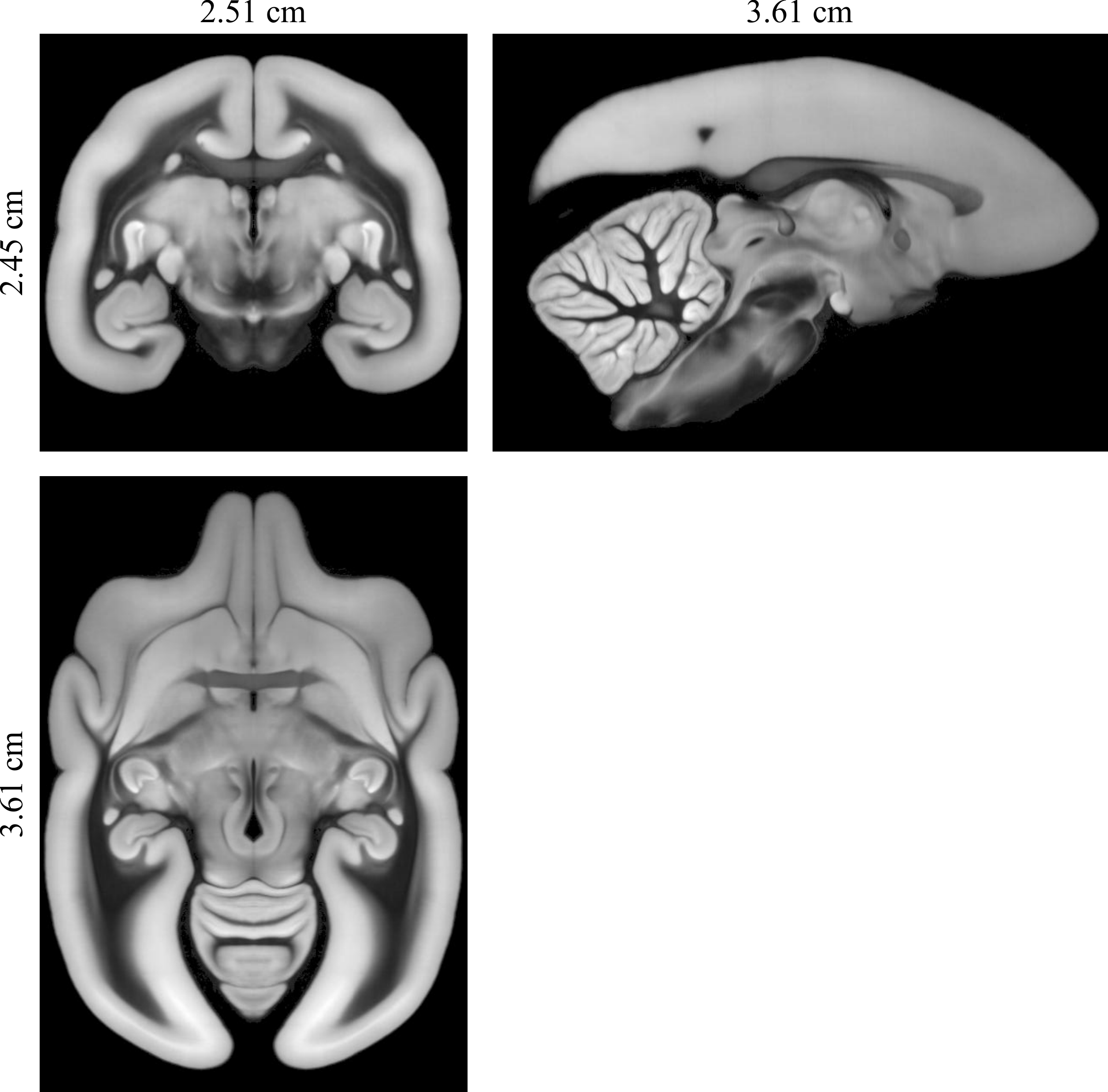}
\caption{The axisymmetric population average image defines the TissueCyte reference image space.}
\label{fig::TCavg}
\end{figure}

We created a TissuCyte average image with a spatial resolution of $50^3 \mu m^3$ based on images of channel $C_R$ from 34 different marmoset monkey brains. We have decided to make the TissuCyte population brain image axisymmetric. The main reason for that was that the number of samples was, at the time the Brain/MINDS program started, even smaller than the 34 images. Compared to the Allen mouse brain atlas template  \cite{kuan2015neuroinformatics,oh2014mesoscale} a rather small number. 
An axisymmetric population average image doubles the number of samples per voxel, which significantly improves the quality of the result. Further, it removes any bias in favor of the left or right hemisphere. Figure \ref{fig::TCavg} shows an ortho view of the TissuCyte population average brain image.

\subsection{Post-Processing}\label{sec::postprocessing}
\setcounter{subsubsection}{4}

\subsubsection{Brain connectivity calculations} After determining injection site location, tracer signal and identifying individual brain regions by mapping the data to the reference brain image space, we can determine the neural connectivity associated with the injection site. 

Therefore, we first identify the connection source area by determining the brain gray matter region(s) that intersect with the injection site. Then, we identify the projection targets by determining gray matter regions that intersect with the axon tracer signal. 
The amount of signal that intersects with the brain regions is used to determine the connection strength between two brain regions. 

Each brain provides complementary information about marmoset brain connectivity. A structural map of marmoset brain connectivity can be obtained by integrating the information of a large number of individual marmoset brains.

\section{Training the U-Net in MarmoNet}

For cell body detection and tracer signal segmentation, in both problems we seek a function $\overline{\mathcal{F}}$ that maps entire images $X:\Omega_\text{in}\to\Rr$ to output images $Y:\Omega_\text{out}\to[0,1]$ with identical domains $\Omega_\text{in}=\Omega_\text{out}$. Each point in $Y(x,y)$ is a label, or a value that is associated with the image content at location $X(x,y)$. In case of the  cell body detection the outputs $Y$ are the cell body saliency maps $\hat{S}^\hi(z)$, in case of the axon tracer signal the functions $\hat{L}^\hi(z)$. 

It is oftentimes challenging, or even impossible to explicitly define the function 
$\overline{\mathcal{F}}$. However, it is frequently possible to define a sufficiently large set of samples $X_i$ from the input domain, and their corresponding samples $\hat{Y}_i$ from the output domain with $\hat{Y}_i=\overline{\mathcal{F}}(X_i)$. Once such a set is given, it is often possible to sufficiently approximate the function $\overline{\mathcal{F}}$ by fitting a trainable function 
$\mathcal{F}\{\xi\}$ to the data, where $\xi$ are the free parameters that need to be determined. The set of pairs of samples $(X_i,\hat{Y}_i)$ is called a training set, because it is used to \textit{train} $\mathcal{F}\{\xi\}$ to approximate $\overline{\mathcal{F}}$.

In our case, the function $\mathcal{F}\{\xi\}$ is an artificial neural network based on the U‐Net network architecture. U-Net is an artificial neural network architecture popular for a broad  variety of image segmentation tasks \cite{falk2019u,ronneberger2015u,cciccek20163d}. U-Net takes images $X:\Omega_\text{in}\to\Rr$ as input and generates output images $Y:\Omega_\text{out}\to[0,1]$.

The creation of the training set for the two different tasks, cell detection and axon signal segmentation, are discussed in more detail in the subsections \ref{sec:apx:inj} and \ref{sec::training::tracer} below. Further technical details regarding the U-Net architecture in this pipeline can be found in section \ref{sec::unet} of the appendix.

\bigskip
\subsubsection{Network Training: Injection site localization}\label{sec:apx:inj}

For cell body detection, we have trained an artificial neural network $\mathcal{F}\{\xi\}$ that maps
full resolution image slices $I^\hi_{{C_B}}(z)$ of channel $C_B$ to cell body saliency maps $\hat{S}^\hi(z)$.

We created a training set by manually labeling the cell body locations of 6068 cells in 44 2D image slices that intersect with the injection site. The images have been selected from 10 \todo{check again} different marmoset brain images. The training images have been divided into smaller, overlapping image tiles that were sufficiently small to be fed into the neural network for training.

\begin{figure}[ht]
\centering
\includegraphics[width = 1 \columnwidth]{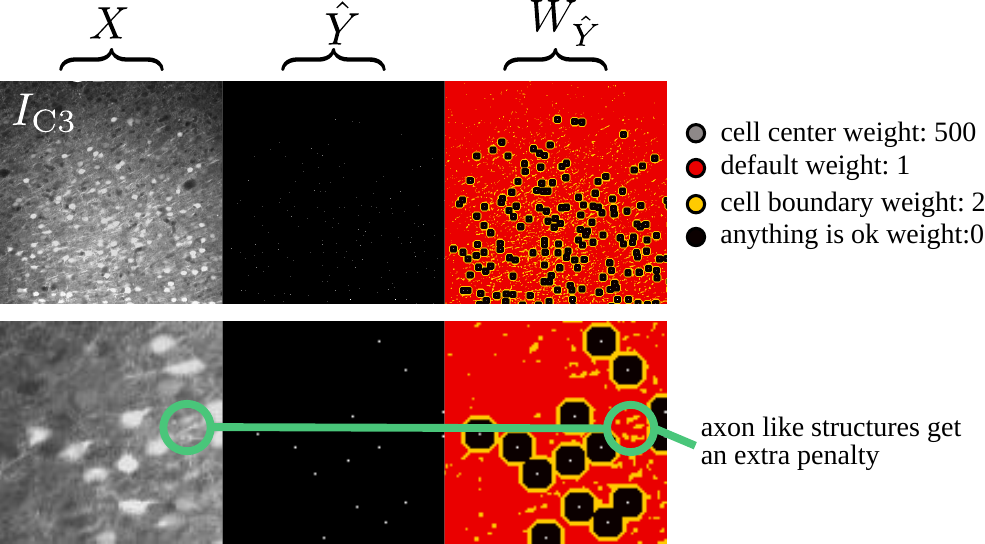}
\caption{An exampled of an image for training a network for cell body detection 
}
\label{Fig::injlocNN}
\end{figure}

Figure \ref{Fig::injlocNN} shows the manual labels of such a training image tile.  The first column shows a training input image $X$. The middle column the ground truth labels $\hat{Y}$. It is a binary image, where only a single pixel at the center of a cell body has been marked positive (=1). All the background has been set to zero. 

It is difficult, error-prone, and often ambitious to manually find and label the exact center of a cell. Further, some bright axons may look similar to cell bodies, and we would like to put an extra penalty to incorrect predictions on those structures during the training. Therefore we automatically generated an additional, spatially dependent weight image $W_{\hat{Y}}$ that scales the error  during the network training based on the image content. The most right column in Fig. \ref{Fig::injlocNN} shows the corresponding example.

The default weight in $W_{\hat{Y}}$ was set to one. In order to tolerate a small displacement between the position of a manually placed label and the position predicted by the network, we masked out any prediction error in a small circular surrounding around a cell body label by setting its weight to zero. To ensure a clear boundary between labels and background, we set the weight of the circle boundaries to two. Further, there are much more negative examples (background pixels) than positive examples (cell body centers). We took this into account by assigning a high weight to the cell body labels (500). Finally, we used a Laplacian of Gaussian filter to detect edge and blob-like structures, thresholded the results and took the result to add an additional weight (=2) to structures, like axons, that may share some similarities with the cell bodies. We determined all parameters heuristically.

\begin{figure}[ht]
\centering
\includegraphics[width = 1 \columnwidth]{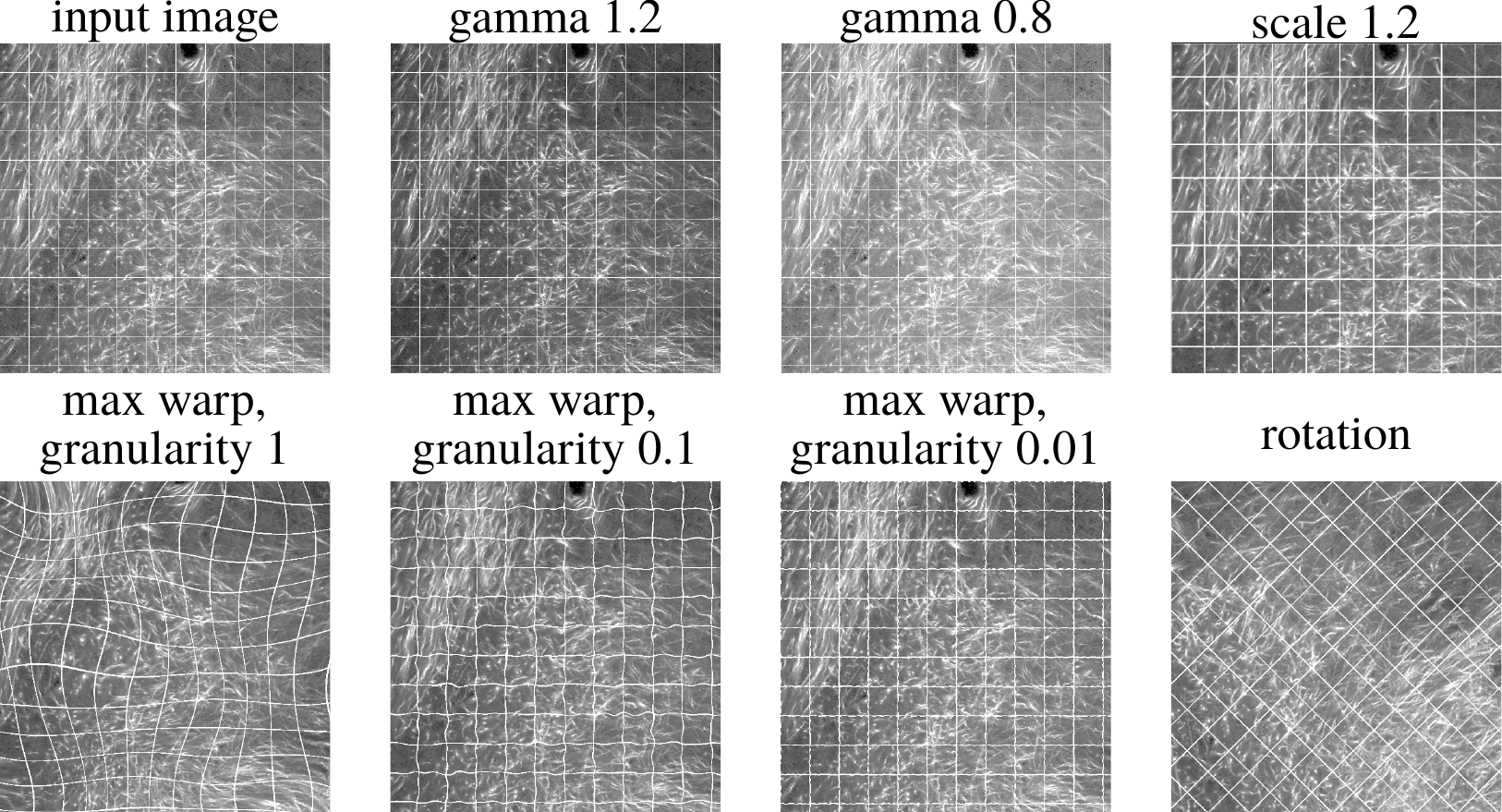}
\caption{Augmentation of training image data. 
}
\label{fig::warping}
\end{figure}

\begin{figure}
\centering
\includegraphics[width = 1 \columnwidth]{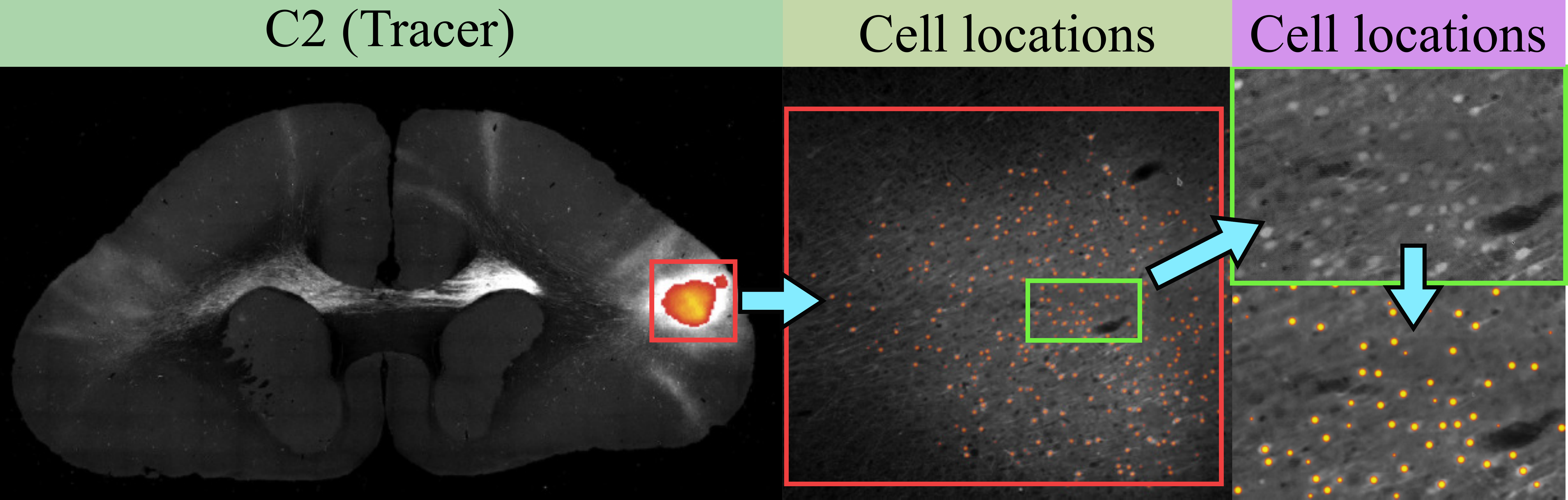}
\caption{The most left figure shows a slice image of the tracer channel $C_G$. As an overlay, the cell density is shown as a heat map at the injection site location. The images on the right show the detected cells.}
\label{Fig::injloc2}
\end{figure}

The samples in the training set may not sufficiently represent all variations in the shape and size of unseen images. 
After training with an insufficient number of samples, the trained network may work well on the training set but then often performs poorly on unseen images. In this situation, data augmentation can help. Data augmentation increases variation in the training set by randomly altering the appearance of the existing images in the training set. In our training, the deformations are one or more combinations of rotations, gamma corrections, smooth, non-linear spatial deformations and global scaling.  Figure \ref{fig::warping} shows examples. An example of detected cells after applying the network to an unseen image is shown in Fig \ref{Fig::injloc2}.

After training with the augmented data, the network was able to segment the axins in all of our 36 different marmosets brain images; see Figure \ref{fig::results01} for the results.

\begin{figure}
\centering
\includegraphics[width = 1 \columnwidth]{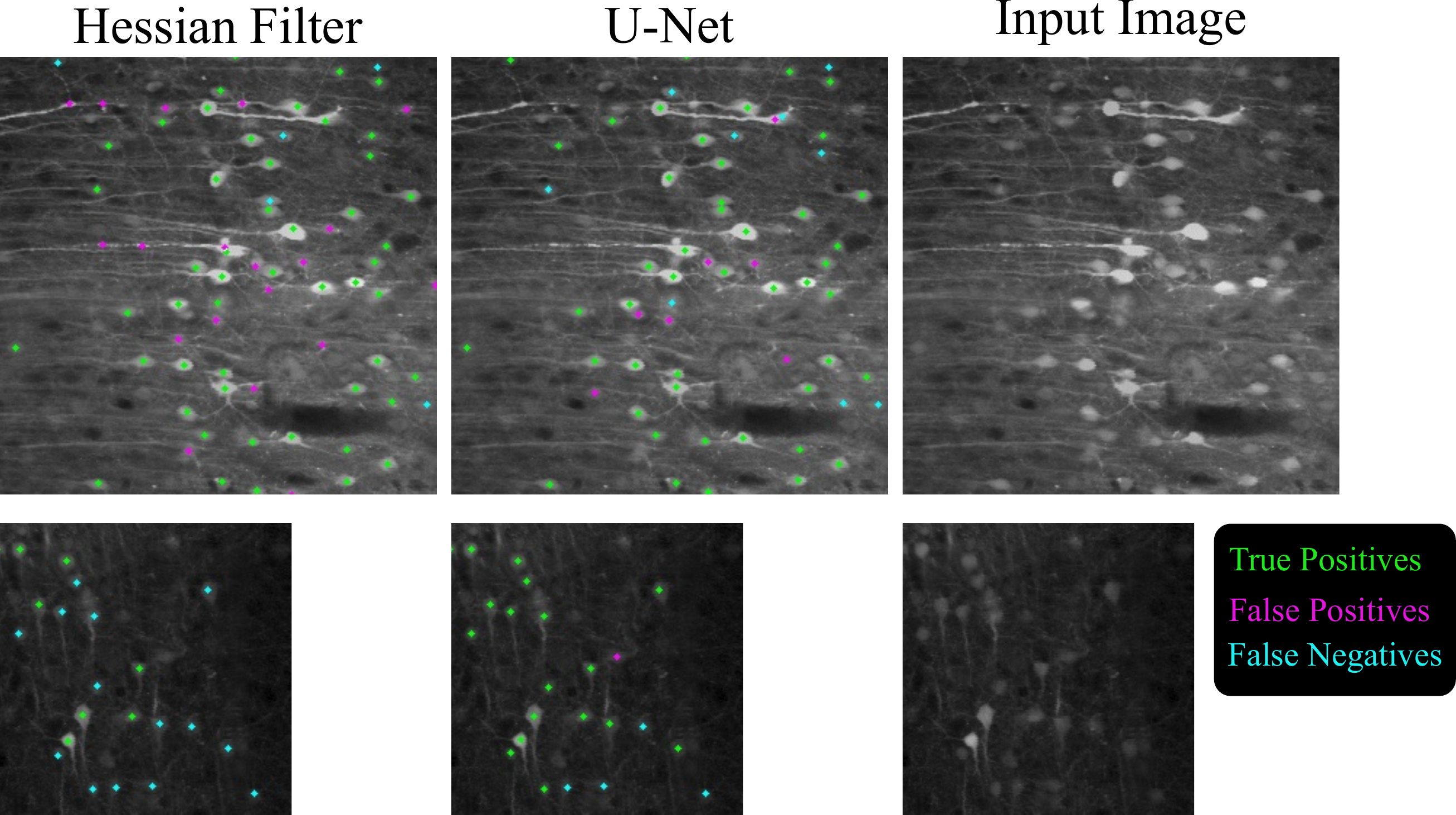}
\caption{Cell body detection using a Hessian filter and the proposed artificial neural network approach. 
}
\label{Fig::PRimg}
\end{figure}

\begin{figure}
\centering
\subfigure[Training data]{\includegraphics[width = 0.44 \columnwidth]{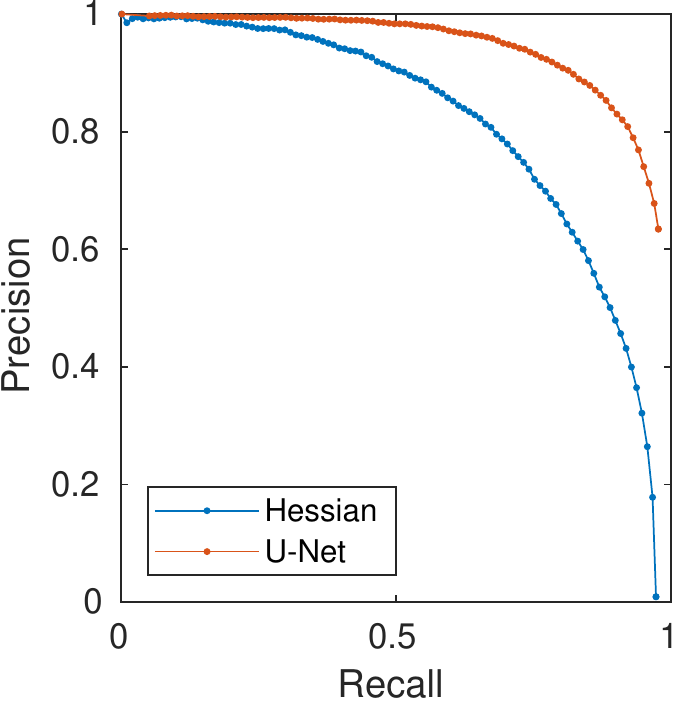}\label{celldet01}}
\subfigure[Test data]{\includegraphics[width = 0.44 \columnwidth]{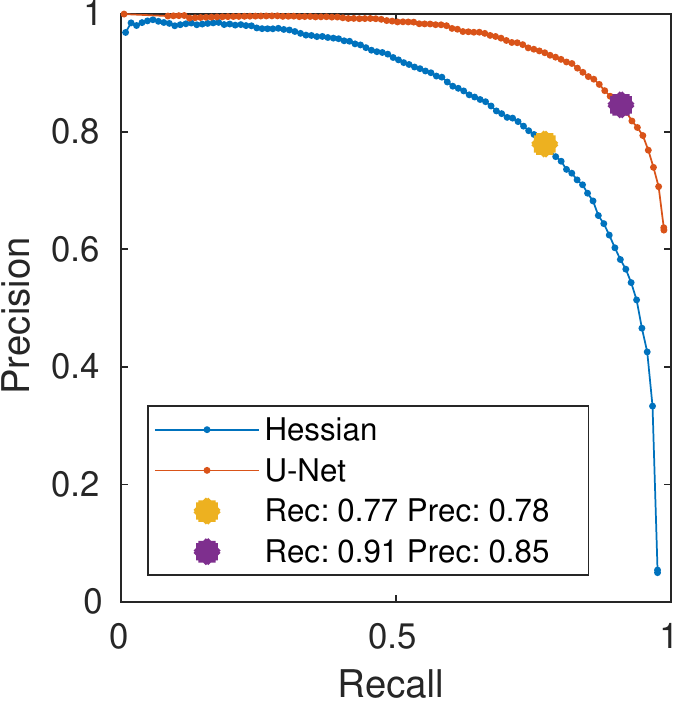}\label{celldet02}}
\subfigure[Performance on the traing dataset (6068 labeled cells in 44 images) and test dataset (3524 labeled cells in 5 images)]{
\centering
\begin{tabular}{l|ccccc}
test dataset& 1 & 2 & 3 & 4 & 5 \\
\hline
labels&1143         &966         &341         &389         &685\\
\multicolumn{6}{c}{ }\\
\end{tabular}
}\label{celldet03}
\caption{Cell detection performance of the U-Net in comparison with a hand-crafted cell detectioon filter.}
\label{Fig::PR}
\end{figure}

\bigskip

\noindent\textbf{EVALUATION:} We have created a test set by manually labeling 3524 cells in five unseen images. Further, we have implemented a hand-crafted cell detection approach and put its detection performance in relation to the neural network approach. Figure \ref{Fig::PR} shows the precision-recall graphs for both approaches when applied to the training set, and when applied to the test set. It clearly shows the superior performance of the neural network approach. Figure \ref{Fig::PRimg} shows a qualitative example.

For the neural network approach, we took the network predictions and detected the local maxima as described in section \ref{sec::injectionsite}. Then we computed precision and recall by varying the threshold $t_\text{high}$ in \eqref{eq::cellmax}.

As an alternative to the neural network, we implemented a blob detection filter capable to detect the roundish shape of the cell bodies within an image. We have implemented a multi-scale filter that analyzes the local curvature of the image based on the second order image derivatives of the Gaussian filtered image. We first applied an isotropic Gaussian filter to the image with standard deviation $\sigma$. Then we computed the Hessian matrix field for the entire image domain and computed the first and second eigenvalues of the matrices. We obtained two new images $\lambda_1$ and $\lambda_2$ of eigenvalues, where, without loss of generality, $|\lambda_1(x,y,\sigma)|>|\lambda_2(x,y,\sigma)|$. With 
\begin{align}
\mathcal{H}(\sigma) = -\lambda_1(\sigma) \frac{|\lambda_2(\sigma)|}{|\lambda_1(\sigma)|+\epsilon} (\lambda_2(\sigma) < 0)
\end{align}
we define the cell detection filter response $\mathcal{H}(\sigma):\Omega\to \Rr$, $\epsilon>0$ is a small constant. The eigenvalues $\lambda_1(x,y,\sigma)$ and $\lambda_2(x,y,\sigma)$ of the Hessian matrix at location $(x,y)$ are proportional to the two local main curvatures within the image plain. 

In case of a bright, roundish plateau, like a cell body, both values are negative. As larger their magnitude, as higher the curvature, as more likely it is a cell. Furthermore, as more roundish the shape is, as more similar should be the two eigenvalues. We took this into account by weighting the term with $-\lambda_1(\sigma)$ (maximum curvature $\rightarrow$ cell contrast), and weight it with the fraction $\frac{|\lambda_2(\sigma)|}{|\lambda_1(\sigma)|+\epsilon}$ (ratio between minimum and maximum curvature $\rightarrow$ cell shape score). The last term $(\lambda_2(\sigma) < 0)$ removes all saddle points such as edges of a cell bodies or axons. 

We have evaluated the performance for all possible combinations of scales (Gaussian filter size) out of the interval $\sigma \in\{2,..,5\}$ on the training set. In case of multiple scales, we took the maximum filter response over all scales. We achieved the best performance with a single scale of $\sigma=4$.

Figure \ref{celldet01} shows the precision-recall graph for the Hessian filter for $\sigma=4$ and the neural network approach on the training set. Figure \ref{celldet02} shows the precision-recall graph on the test set. We have marked the threshold that performed best (precision equals recall) during training in the test graph indicating that both approaches can maintain their performance on the test set.

\bigskip

\subsubsection{Network Training: Axon tracer signal detection}\label{sec::training::tracer}

\begin{figure}[ht]
\centering
\includegraphics[width = 1 \columnwidth]{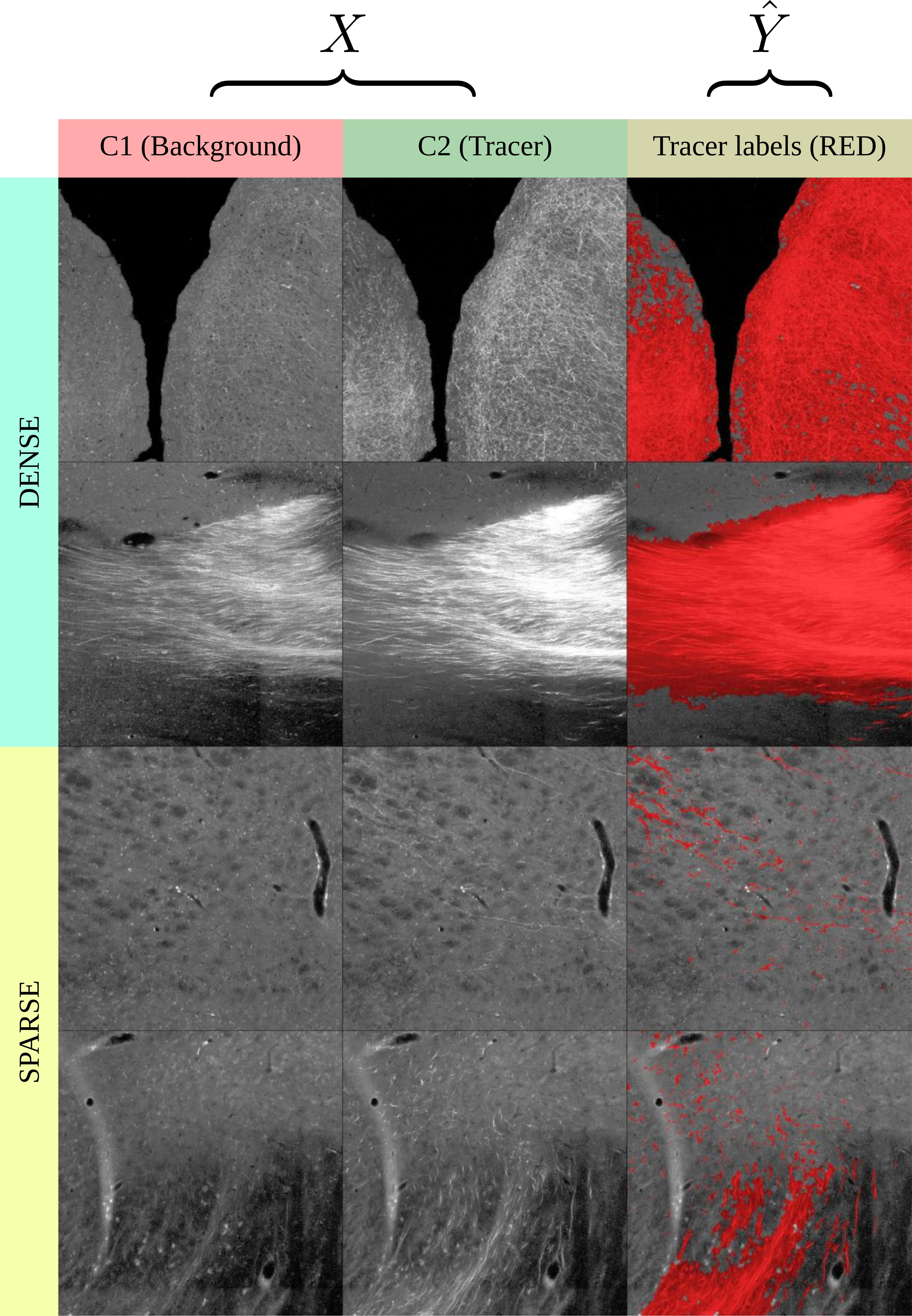}
\caption{The threshold based approach worked well for many images. We manually selected the best candidates for training the neural network.}
\label{Fig::training:pos}
\end{figure}

\begin{figure}[ht]
\centering
\includegraphics[width = 1 \columnwidth]{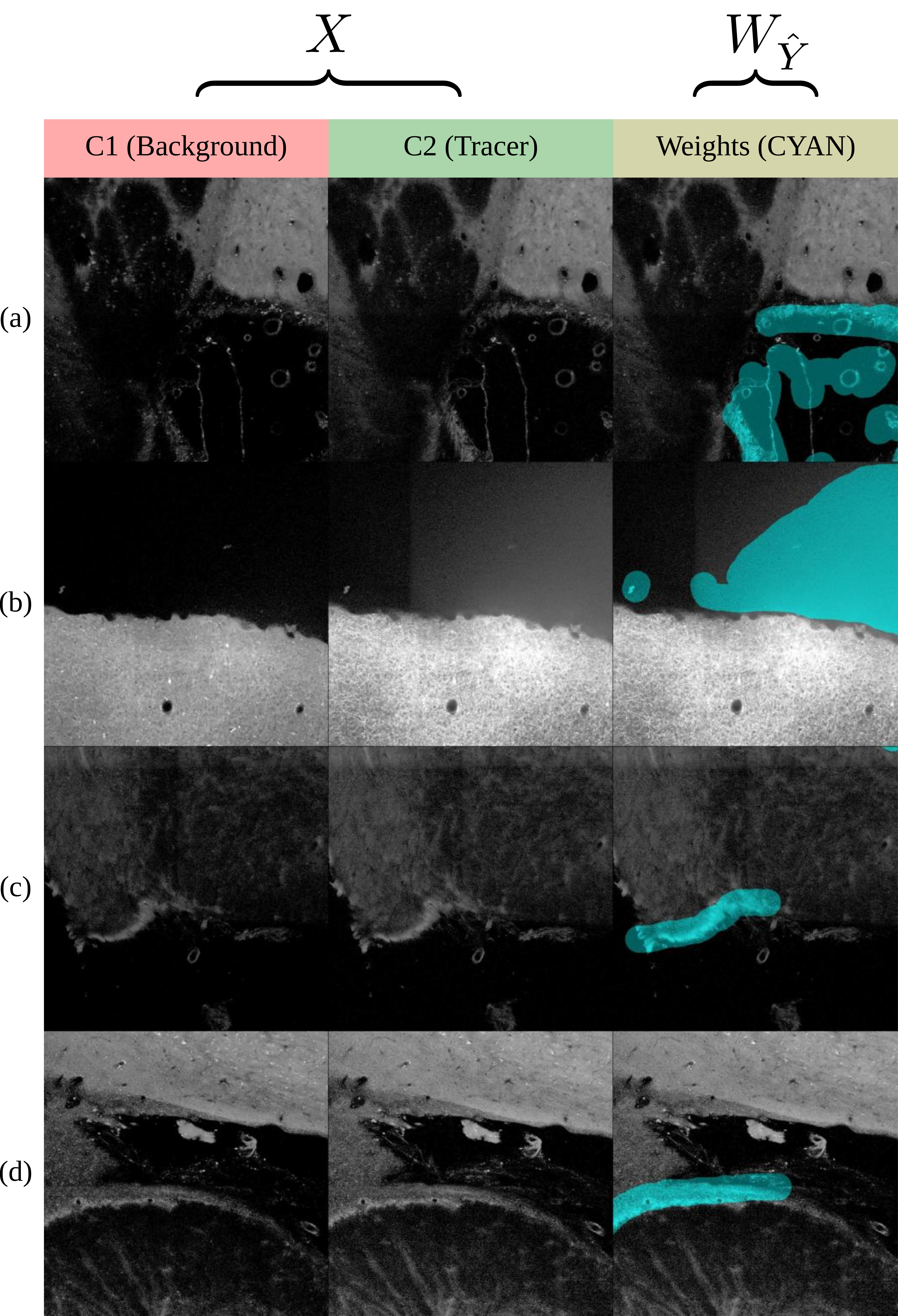}
\caption{Bright structures, like blood vessels, frequently lead to false positive detections. We manually labeled many of them as negative examples for training the U-Net.}
\label{Fig::training:neg}
\end{figure}

For axon tracer signal detection, we have trained an artificial neural network $\mathcal{F}\{\xi\}$ that takes two full resolution image slices $I^\hi_{{C_R}}(z)$ and $I^\hi_{{C_G}}(z)$ of channel $C_R$ and channel $C_G$ as input and maps them to the axon tracer saliency map $\hat{L}^\hi(z)$.  

In contrast to the cell body detection task, it was impractical to manually annotate a sufficiently large amount of images for training the network for the axon tracer detection.  The axons appear in various shapes and with varying contrasts, intensities and densities. Some axons appear as elongated, bright line-like structures with clear boundaries where a manual annotation may be feasible. However, many others build dense, blob-like or cluttered patterns with a texture that can vary depending on whether they cross and bifurcate or they share a dominant direction as part of a thicker axon bundle.

Instead of manual annotation, we used an approach based on intensity thresholding to obtain a large set of labeled brain image sections and afterwards manually selected the best candidates for training. We selected about 600 image sections from 20 different marmoset brain image stacks. Even within the 600 image slices, various non-axonal structures, like the boundary of some blood vessels, lead to false-positive detections. After selecting the 600 best candidates, we manually annotated such structures in the images to provide some explicit negative examples to the network. Figures \ref{Fig::training:pos} and \ref{Fig::training:neg} show examples of the thresholded tracer signal, and manually annotated background structures, respectively. 

From the 600 labeled full-brain image slices, we automatically generated a training set containing about 12000 smaller image tiles: We smoothed the labels (the thresholded tracer signal) with a large isotropic Gaussian function to obtain a tracer signal density map. We used the density map to generate two probabilistic density functions to randomly pick 20 smaller image tiles from each of the 600 full resolution image sections. The first function was designed to draw 10 tiles from axon dense regions while the second one was used to draw 10 tiles from axon sparse regions.  About one-third of the full brain slices did not contain any tracer signal. When there was no tracer signal in a full brain slice then all 20 image tiles that were generated from that slice were only background signal (about 1/3 of the data). Further, all structures with manually labeled negative examples have been divided into smaller tiles and have been added to the training set. The size of an image tile was $810 \times 810$ pixels which then has been cropped, after augmentation, down to the network input size of $572 \times 572$ pixels.

The ratio of background to tracer signal in the training data was about eight to one. To balance the different classes, we weighted the error of incorrectly predicted tracer labels by eight. The weights for manually annotated negative examples have been set to 100.


\bigskip
\noindent\textbf{THRESHOLDING:}  The axon tracer signal is most dominant in the green color channel $C_G$; see Fig. \ref{Fig:microscope}. While it is significantly stronger in $C_G$ than in $C_R$, the autofluorescent background signal is mostly identical in both channel (up to a small constant factor).

\begin{figure}[ht]
\centering
\includegraphics[width=1\columnwidth,height=\textheight,keepaspectratio]{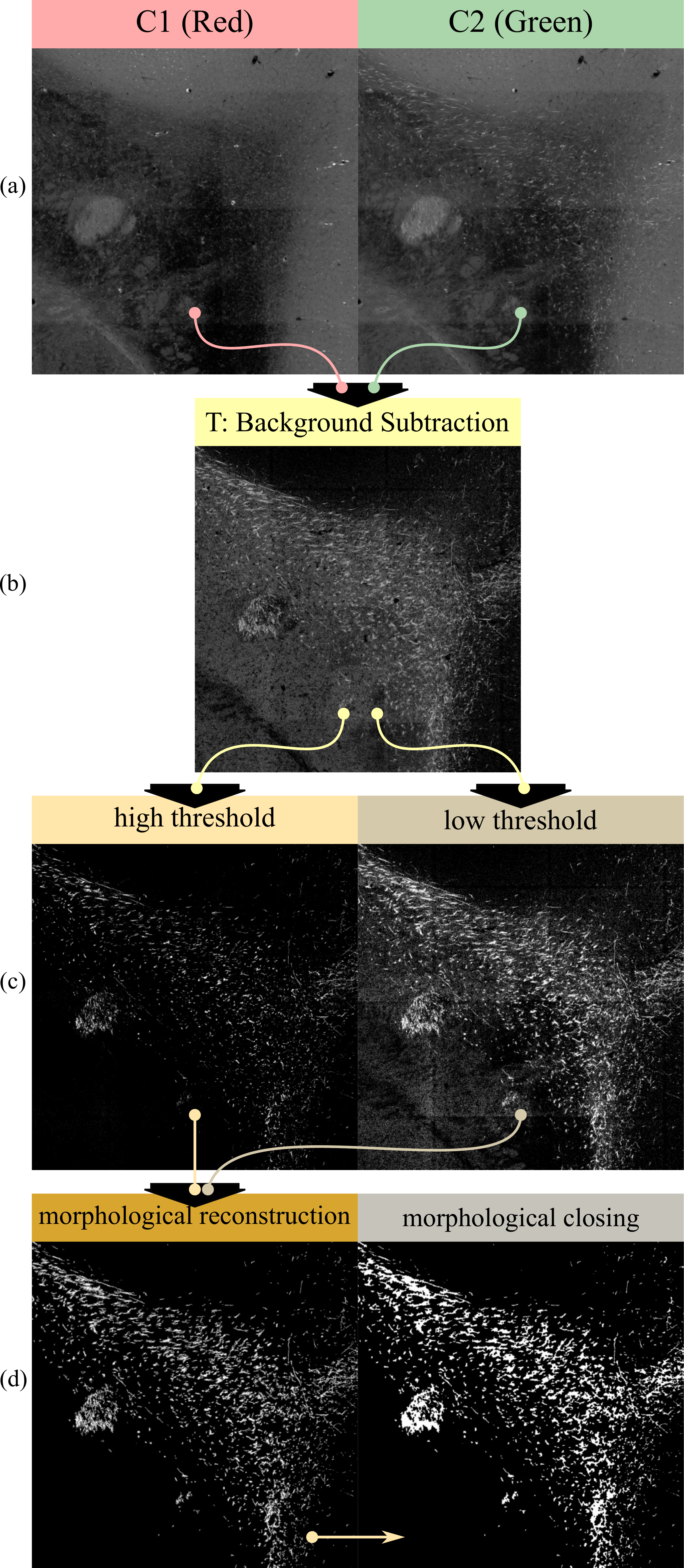}
\caption{Morphological pipeline for creating axonal tracer training data}
\label{Fig:threshold}
\end{figure}

Subtracting channel $C_R$ from $C_G$ increases the contrast between the tracer signal and the background; see Fig. \ref{Fig:threshold} (b). With equation \eqref{eq::tracerthreshold}, we can apply a threshold to that high-contrast image to obtain a tracer signal segmentation mask. However, determining a good threshold that separates the background from the tracer signal is tricky. The varying brightness and contrast of the axon tracer signal within different brain regions make the determination of a good threshold value difficult. It is always a trade-off between a low false-negative rate or a low false-positive rate. Automated methods like Otsu's method  \cite{otsu1979threshold} did not perform well.

We obtain good results by using a double-threshold approach. A high threshold separates the bright tracer signal from the background. The high threshold has been chosen conservatively so that we were confident that no background structures remains in the image. A second, lower threshold has been determined that recovers as much tracer signal as possible while keeping the false positive detections low. We did a grid search over threshold parameters and have qualitatively compared the results by visual inspection. We determined 300 for the high threshold, and 100 for the low threshold. In Fig. \ref{Fig:threshold} (c) we show examples. 

The lower threshold retrieves the weak tracer signal thas has been classified as background by the higher threshold. However, it incorrectly classifies many other small structures as tracer signal as well. In the next step, we use morphological reconstruction to select those connected regions out of the low-threshold image that have - at least - one pixel in common with the high threshold image. The idea is that the high threshold image selects the reliable regions in the low threshold image. The first image in Fig. \ref{Fig:threshold} (d) shows an example. In a final step, we use morphological closing with a disk-shaped structure element of size three to smoothen the result.

\begin{figure}[ht]
\centering
\includegraphics[width = 1 \columnwidth]{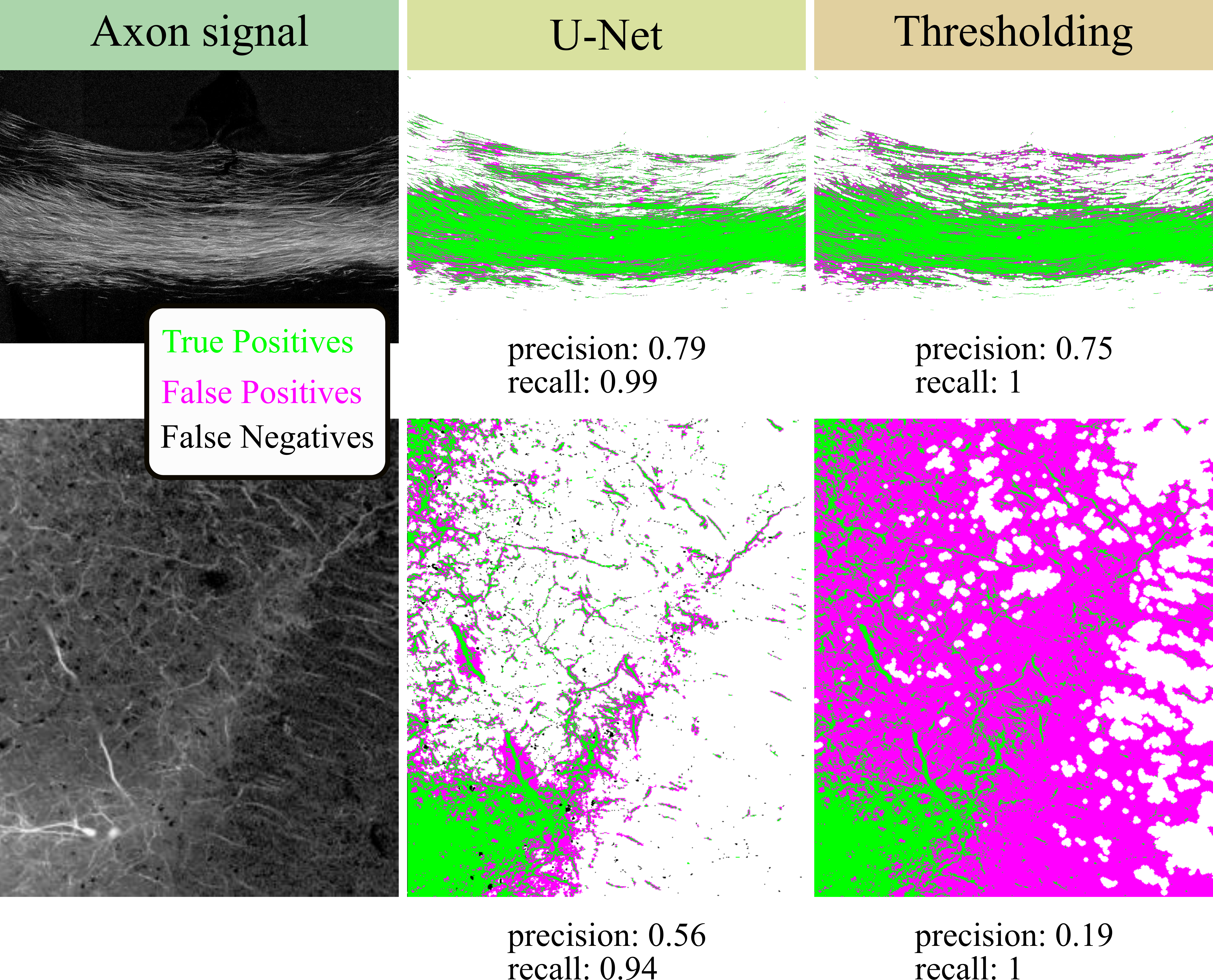}
\caption{Axon tracer signal detection. 
}
\label{Fig::axoneval}
\end{figure}

\bigskip
\noindent\textbf{EVALUATION:} We exemplary selected two unseen images with axon tracer signal. One image with densely labeled axons that share the same main direction as part of a thicker axon bundle in the corpus callosum. The other image from branching axons with varying density from within the brain gray matter. For both images, we subtracted channel $C_R$ from channel $C_G$ and manually tuned the image intensities for both images to increase the contrast between the axon signal and the background signal. Then, we applied a threshold to obtain a segmentation mask and manually corrected errors. We took the manually segmented image as ground truth and compared it with the two automated approaches, the thresholded output of the neural network, and the output of the double-threshold approach. Figure \ref{Fig:threshold} shows the results. The last column illustrates the difficulty of finding a threshold parameter that works well for all images. While performing well for the image of the axon bundle, it is too sensitive for the other image leading to many false positive detections. The neural network approach has a clear advantage. Data augmentation provides an automated method to prepares the network for unseen images with a contrast or  brightness that is different to the images in the training data.

\section{Discussion \& Conclusion}
We are developing a pipeline that processes, analyzes and maps tracer image data to a common marmoset brain image space. The pipeline incorporates state-of-the-art machine learning techniques based on artificial convolutional neural networks (CNN) and state-of-the-art image registration techniques to extract and map all relevant information in a robust manner. The pipeline processes images in a fully automated way. 

In Fig. \ref{fig::results01} we show results of the tracer segmentation and the corresponding cell densities in the injection site. All images have been mapped to the Brain/MINDS marmoset brain atlas reference image space.

In this report, we briefly introduced all steps of the tracer signal processing pipeline. The two major components of the image analysis part of the pipeline have been explained in more detail, which are: the injection site detection and the axon tracer signal segmentation.



In future versions of the image analysis pipeline, we plan to improve the axon tracer signal segmentation. Axon tracer signal segmentation is difficult due to the broad variety of shape and contrast of the axon signal pattern, and the density of labels. Completely manually creating a sufficiently large ground truth database for training a CNN remains impractical. However, some kind of interactive relevance feedback system might be appropriated, where an artificial neural network is capable to automatically detect ambiguous classification results and gets feedback from human experts to improve its performance. With such an approach we may be able to create a smaller and better performing training set with reasonable effort and time.

It is often impractical, error-prone, and/or strongly biased to analyze a large amount of data in a manual manner so that automated approaches are strongly demanded. Of course, the lack of correct ground truth data makes it difficult to perfectly master all those problems with automation. Nevertheless, automated processing is often the only option to gain new insights from large scale imagining data. We believe that the best practice to evaluate the data and the results is to share the results with a broader community of experts by making them publicly available.  We plan to make our results publicly available in the near future.

\newpage

\def\thesubsubsectiondis{(\arabic{subsubsection})} 
\appendix

\subsection{Flat-field correction}\label{sec::flatfieldcorr}

\begin{figure}[ht]
\centering
\subfigure[before]{
\includegraphics[width = 0.47 \columnwidth]{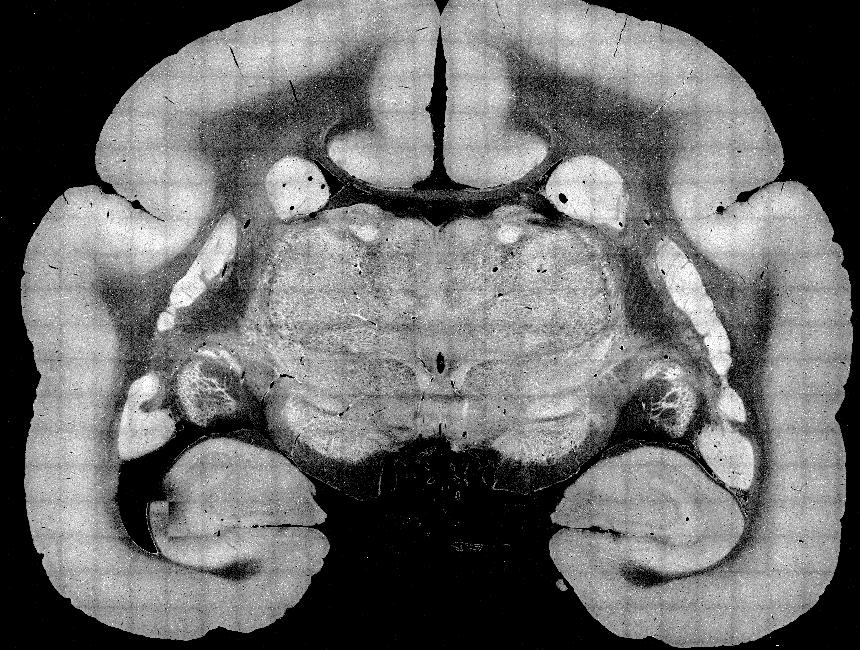}}
\subfigure[after]{
\includegraphics[width = 0.47 \columnwidth]{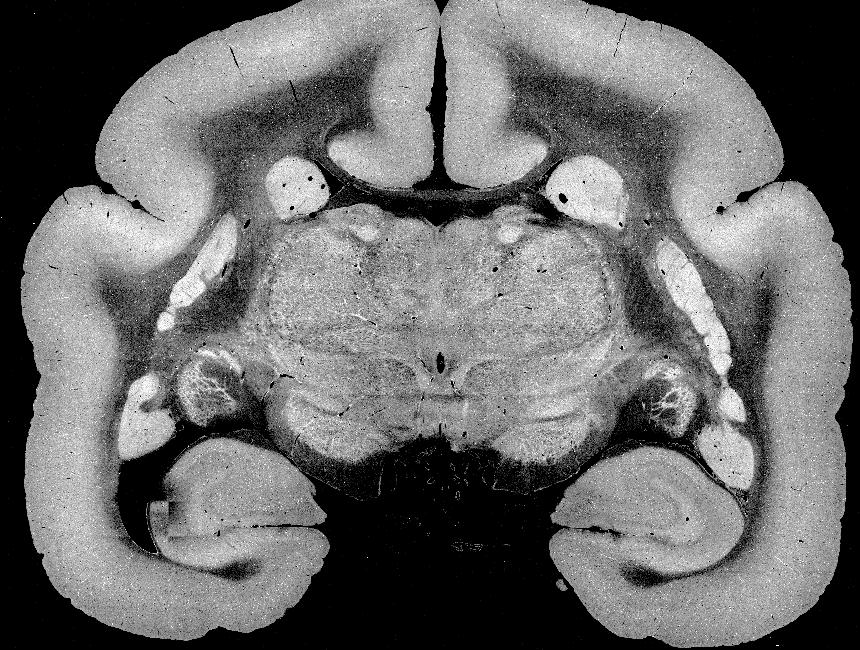}}
\subfigure[before]{
\includegraphics[width = 0.47 \columnwidth]{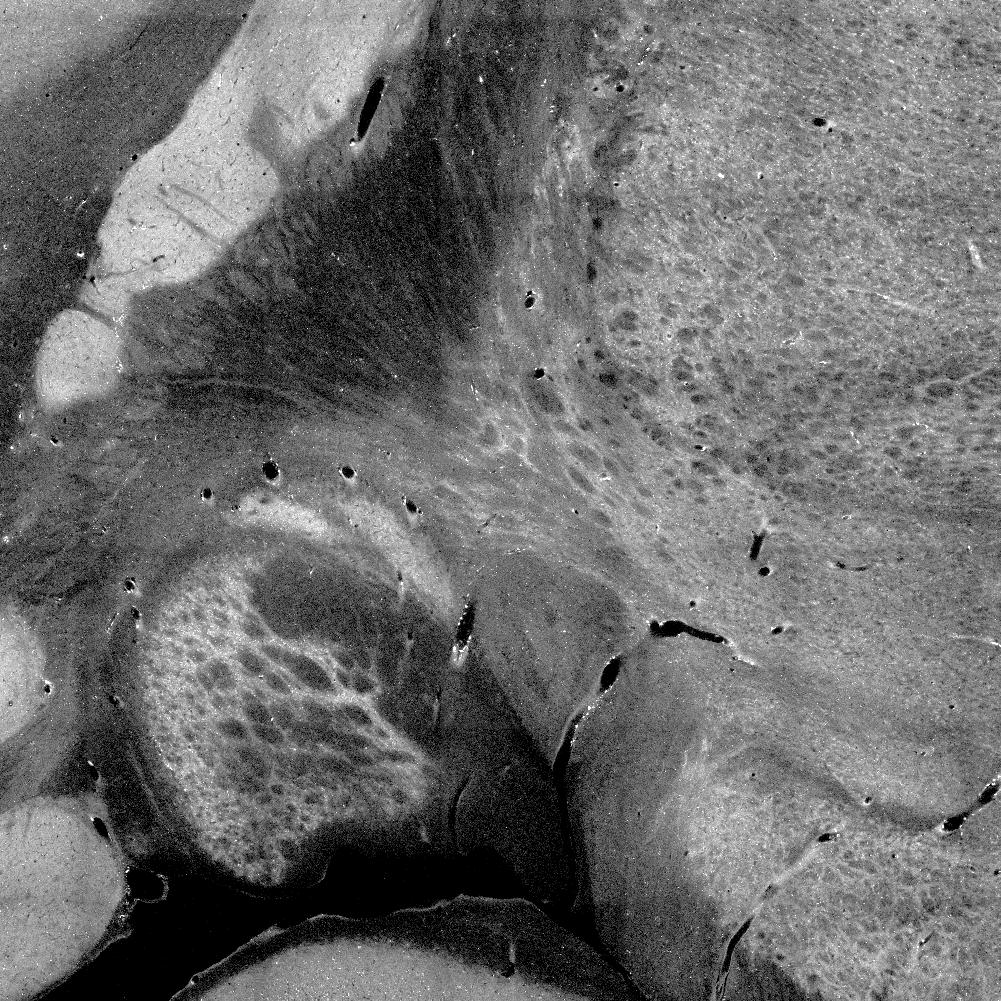}}
\subfigure[after]{
\includegraphics[width = 0.47 \columnwidth]{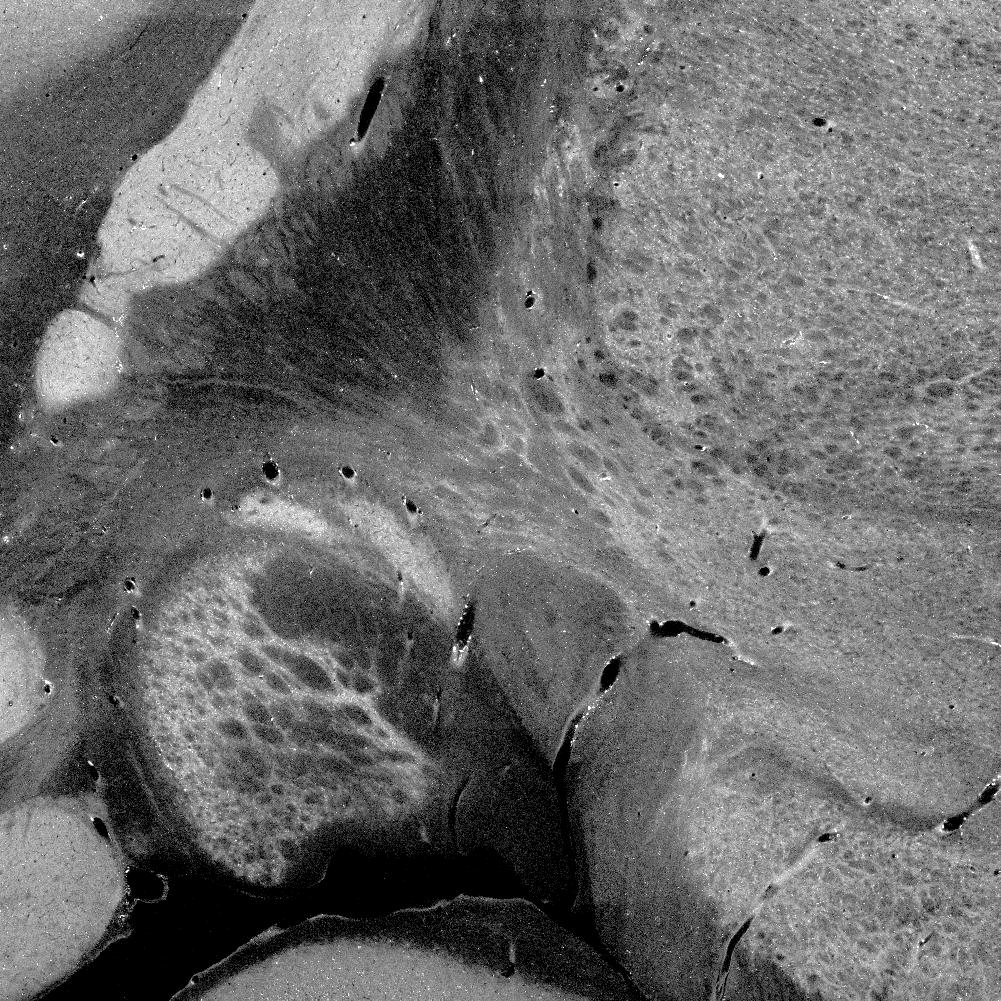}}
\caption{The flat-field correction noticeably reduces the vignetting effect. 
}
\label{Fig::FFC02}
\end{figure}

The relation between the true image $\org{I}$ and the observed image $I$ is often modeled by  
\begin{align}
I = \org{I} \cdot F + D;
\label{eq::intcoo01}
\end{align}
see for instance \cite{kask2016flat,likar2000retrospective}.
 The function $I:\Omega \to \Rr_{\geq 0}$  is the image tile, $D:\Omega \to \Rr_{\geq 0}$ the darkfield, and $F:\Omega \to \Rr_{\geq 0}$  a multiplicative shading component which actually causes the vignette effect. With $\Omega$ we denote the image domain. The darkfield $D$ represents the amount of signal given out by the detector when there is no incident light. A flat field can be corrected by inverting the model. The inverting is defined by
\begin{align}
\org{I} = \frac{I - D}{F}.
\end{align}
Removing the bias $D$ is often necessary to calibrate CCD sensor based cameras and microscope systems. For our two-photon microscope, however, which scans the probe in a raster pattern pixel-by-pixel, it can be neglected. The flat-field correction then simplifies to 
\begin{align}
\org{I} = \frac{I}{F}.
\label{eq::intcoo012}
\end{align}

The vignetting effect slightly differs from microscope to microscopes such as for our TC1000 and our TC1100. The different pathways of the different color channels cause different effects as well. 

Similar to \cite{allentecrep,abdeladim2019multicolor,guesmi2018dual}, we use a data-driven heuristic to estimate the shading field $F$. When averaging over a  sufficiently large number of image tiles $I^j$ we obtain an estimate of the shading field image $F$ (up to a constant factor).

With \eqref{eq::intcoo012}, we have the relation
\begin{align}
\frac{1}{M}\sum_{i=0}^M {I^j} = F  \frac{1}{M}\sum_{i=0}^M \org{I}^j.
\label{eq::intcoo04}
\end{align}
We assume that all pixel intensities in the tiles $\org{I^j}$ are Poisson distributed in the same manner with mean value $\lambda  \in \Rr_{>0}$. The averaging over a large number of sample tiles $\org{I^j}$   leads to a flat mean intensity image with a constant value close to $\lambda$. In this case, equation \eqref{eq::intcoo04} simplifies to 
\begin{align}
 F \underset{\approx \lambda}{\underbrace{\frac{1}{M}\sum_{i=0}^M \org{I}^j}} = F \cdot \lambda.
\label{eq::intcoo05}
\end{align}
Let now 
\begin{align}
\hat{F}:=\frac{F \cdot \lambda |\Omega|}{\sum_{\vec x \in\Omega}(F(\vec x) \cdot \lambda)}
 \end{align}
be our normalized shading field approximation. Then we can approximate the flat-field corrected image $\org{I}^j$ with
\begin{align}
\org{I}^j \approx  \frac{I^j}{\hat{F}}.
\end{align}

An entire marmoset brain image contains a large number of tiles. We maintain a low memory footprint by using a running average to approximate $\hat{F}$. Further, we exclude outliers by discarding all dark pixels with an intensity lower than 2, and all bright pixels with an intensities greater than 2500 (axon tracer signal) during the average computation.

\subsection{U-Net}\label{sec::unet}

\begin{figure}
\includegraphics[width = 1 \columnwidth]{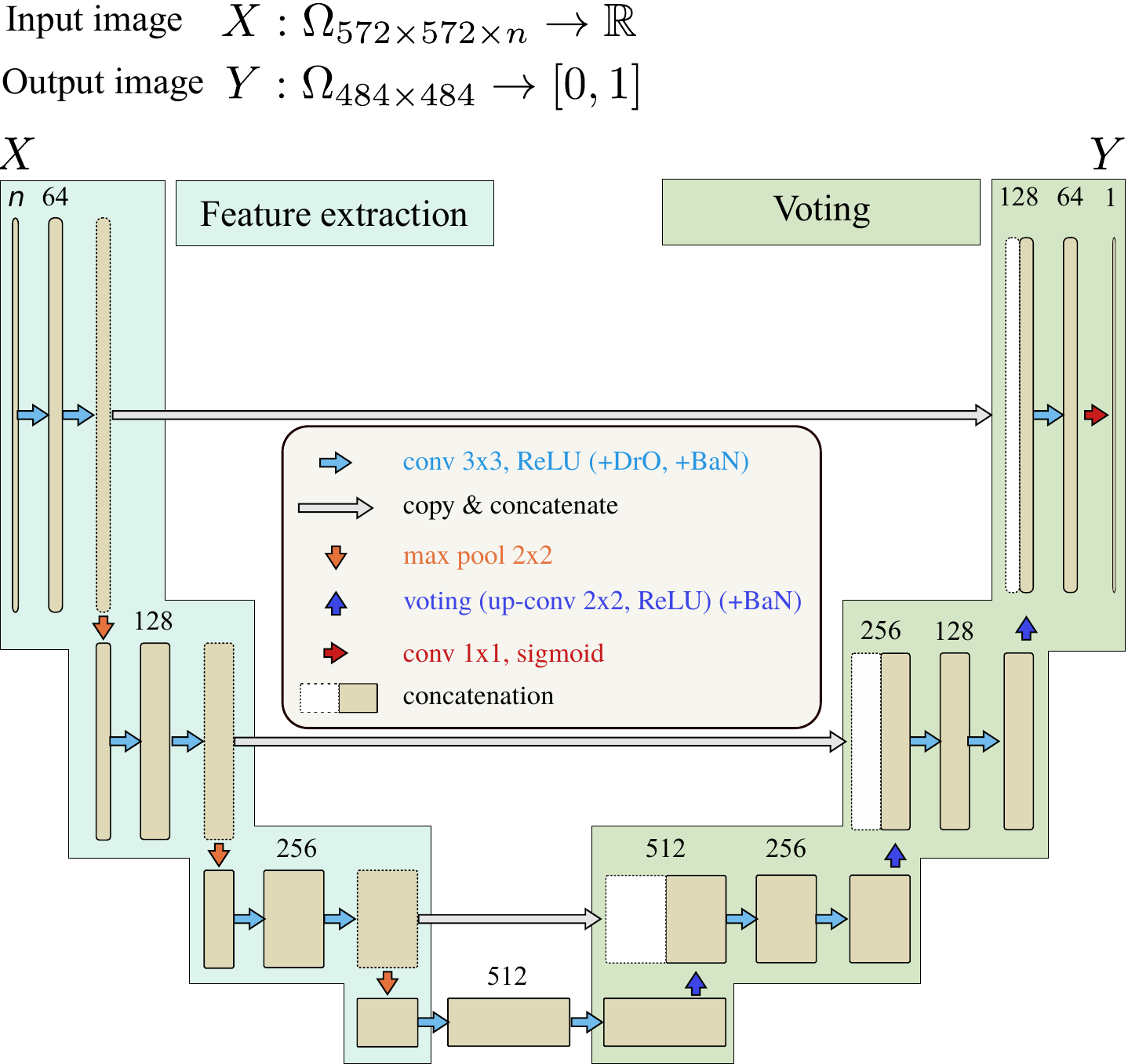}
\caption{The MarmoNet U-Net network architecture for cell body detection and tracer signal segmentation.}
\label{Fig::Unet}
\end{figure}

A U‐Net consists of a feature extraction path and a voting path; see the left and right part of Fig. \ref{Fig::Unet}. With $n$ we denote the number of channels of the input image. For the cell body detection problem, where only $C_B$ is fed into the network, it is $n=1$. For the tracer segmentation tasks, $C_R$ and $C_G$ are being concatenated so that the input dimension is $n=2$ (two values per pixel).

The feature extraction path is a standard convolutional neural network. It consists of repeated applications of unpadded $3\times3$ convolutions followed by rectified linear units (light blue arrows) and $2\times2$ max-pooling (orange arrows). The free parameters are the convolution kernels. The objective of the network training is to determine the kernel weights that extract the image features from the input image that are best suited to solve the task.

Starting with 64 features after the first convolution layer, the number of features is doubled after each pooling operation to up to 512 features. Each pooling operation halves the extents of a feature image by taking the maximum over a four-pixel neighborhood. The pooling is an efficient way to enlarge the perceptive field of the neural network while keeping the computation costs on a reasonable level.

The voting consists of repeated applications of $2\times2$ up-convolutions (dark blue arrows) followed by applications of $3\times3$ convolutions and rectified linear units. Each up-convolution operation doubles the extent of a feature image. Starting with 512 features, we  halve the number of features after each up-convolution down to 64. 

While in the feature extraction path, the convolutions can be considered as an information extraction operation, the convolutions in the voting path can be considered as a drawing operation. During voting, each convolution kernel can be considered as a basis function that locally draws a specific pattern into the output image $Y$. The basis functions, determined by the kernel's free parameters, are tuned during the training. However, the weights that determine the contribution of each basis function in the image generation process are determined by the extracted features in the feature extraction path.  The superposition of all those basis functions generates the final output.

The last activation function in the network is a sigmoid function that generates the label image $Y$ by mapping all values to the interval $[0,1]$.

U-Net employs direct intermediate connections from feature layers to voting layers (gray connections in Fig. \ref{Fig::Unet}). The connections propagate spatial features directly from
high‐resolution feature layers to deeper voting layers. Those features support a precise spatial localization of the output labels.

We further use drop out and batch normalization \cite{ioffe2015batch,srivastava2014dropout} to improve the training performance; see Fig. \ref{Fig::Unet}, where we have labeled dropout layers with \textit{DrO} and batch normalization layers with \textit{BaN}.

The image patch input size of our network is $572\times572\times n$ pixels. The output image is, due to the unpadded convolutions, cropped down to $484\times484$ pixels. Note that since all operations are convolutions, the image size can be changed during, or even after the training. For training, however, $572\times572\times n$  was a good trade-off regarding GPU memory size and computation time.

When it comes to applications, an entire input image is rather large, and the amount of memory of a modern graphics card, which is used to perform the network processing, is limited. When applying the network to large image slices, a sliding window approach is used to move the receptive field of the network over the entire image to label its content one after another. This is done by dividing a large image into smaller image tiles,  feed them into the network one after another, and stitch the network outputs together to form the entire output image.

For energy minimization during training, we used weighted logistic regression as a loss function.

A problem that appears quite often with biomedical image data is the small number of training samples. 
The number of free parameters in U-Net is large, and the samples in the training set may not sufficiently represent all variations in the shape and size of unseen images. As a result, often, the trained network works well on the training set but then performs poorly on unseen images. This is called overfitting. We increased the variation in the training set by altering the appearance of the existing images in the training set randomly. Figure \ref{fig::warping} shows examples. The deformations are combinations of one or more transformations. We applied rotations, gamma corrections, smooth, non-linear spatial deformations and global scaling. The process is known as image augmentation.

%

\end{document}